\documentclass[prd,aps,superscriptaddress,amsmath,amssymb,amsfonts,twocolumn,floatfix]{revtex4-2}
\usepackage{amssymb,amsmath,amsthm,amsbsy,color,graphicx,times}
\pdfoutput=1
\usepackage[ansinew]{inputenc}
\usepackage[english]{babel}
\usepackage{color}
\usepackage{braket}
\usepackage{booktabs}
\usepackage{tabularx}
\usepackage{array}
\usepackage{bbold}
\usepackage{natbib}

\begin{document}

\title{Beyond the MSW effect: Neutrinos in a dense medium.}

\author{A. Capolupo}
\email{capolupo@sa.infn.it}
\affiliation{Dipartimento di Fisica ``E.R. Caianiello'' Universit\`{a} di Salerno, and INFN --- Gruppo Collegato di Salerno, Via Giovanni Paolo II, 132, 84084 Fisciano (SA), Italy}

\author{S. M. Giampaolo}
\email{sgiampa@irb.hr}
\affiliation{Institut Ru\dj er Bo\v{s}kovi\'c, Bijeni\v{c}ka cesta 54, 10000 Zagreb, Croatia}

\author{A. Quaranta}
\email{anquaranta@unisa.it}
\affiliation{Dipartimento di Fisica ``E.R. Caianiello'' Universit\`{a} di Salerno, and INFN --- Gruppo Collegato di Salerno, Via Giovanni Paolo II, 132, 84084 Fisciano (SA), Italy}

\begin{abstract}

We present a theory of neutrino oscillations in a dense medium which goes beyond the effective matter potential used in the description of the MSW effect.
We show how the purity of the neutrino state is degraded by neutrino interactions with the environment and how neutrino--matter interactions can be a source of decoherence.
We present new oscillation formulae for neutrinos interacting  with leptons and carry out a numerical analysis which exhibits deviations from the MSW formulae for propagation through the Earth of ultra-high energy neutrinos.
In particular, we show that at high density and/or high neutrino energy, the vanishing transition probabilities derived for MSW effect,  are non zero  when the scattering is taken into account.

\end{abstract}

\maketitle

\section{Introduction}
\label{sec.Introduction}

Neutrinos are notoriously elusive particles with an extremely low interaction rate~\cite{Vannucci}, which oscillate among three different flavors~\cite{Nakamura1, Nakamura2, Nakamura3, Nakamura4, Nakamura5, Nakamura6, Pontecorvo, Capolupo2018}.
They only participate in weak and gravitational interactions, but the latter can be neglected, for all practical purposes, in virtue of their small mass~\cite{Nakamura6} ( except for very peculiar situations~\cite{CapolupoCPTdiss}).
Indeed, the primary modification induced by matter on neutrino oscillations, known as the MSW effect~\cite{MSW1, MSW2} is entirely due to the weak interactions of neutrinos with the electrons and nucleons of the environment.
The MSW effect is, in essence, an effective theory, since all the interactions are comprised in an effective position--dependent potential  acting differently on the individual neutrino flavors and hence affecting the oscillation probabilities.
 However, there is one major shortcoming in adopting such an effective field approach to neutrino--matter interactions, and it is the complete loss of insight on the dissipative aspects of these interactions.
 Indeed, the unitary evolution driven by the effective potentials $V_{\alpha} (\pmb{x})$  always preserves the purity of neutrino states.
While this is reasonable for neutrino oscillations  in a low-density medium, for the interactions in a  denser one, a loss of coherence due to collisions with the  particles in the medium is to be expected.

 Decoherence effects in neutrino oscillations have been thoroughly studied from both a phenomenological and a theoretical standpoint.
To include such effects, the neutrino is usually modeled as an open system in which the action of a dissipative dynamics is discussed on quite general grounds~\cite{Lind,Benatti, Benatti1}.
Exploiting this approach it has been shown that the decoherence leads to modified oscillation formulae~\cite{Benatti} that can shed some light on the fundamental nature of neutrinos~\cite{CapolupoDec}. The main problem is that the dissipative dynamics of neutrinos depends  on several external parameters, that are not determined within the theory, hence making the comparison with the experiments extremely hard.

Here, we present a novel approach to neutrino oscillations in matter, in which the loss of coherence due to the collisions with the particles in the medium is explicitly taken into account. We consider oscillations in a dense medium, i.e. in a medium where the electronic density strongly suppresses the flavor oscillations
and we provide a theory
 which goes beyond the effective matter potential, and give a quantitative account of the quantum decoherence induced by the interactions with the environment.
To this end, one needs to resolve the details of the single scattering events and estimate the loss of coherence due to the latter. Therefore, our considerations are in principle valid for the propagation of neutrinos in any environment, regardless of the frequency of collisions.
This includes also the regimes in which, actually, the neutrino evolution is analyzed in terms of effective damping of the oscillations~\cite{Stodolsky1987}, by means of a Boltzmann--like collision integral~\cite{Raffelt1993} and through a kinetic equation for neutrinos~\cite{Raffelt2018,Vlasenko2013, Blaschke2016}.

We show how the purity of the neutrino state is degraded as an effect of its interactions with the environment that can be interpreted as the quantum decoherence arising from a dissipative Markovian evolution.
 This provides a microscopic qualitative and quantitative explanation to the phenomenology of decoherent neutrino oscillations as analyzed in previous works.
Such a  phenomenon can be of particular interest, not only for a deeper understanding of neutrino oscillations in the matter but also for its phenomenological implications.
 In this picture, we obtain new oscillation formulae with respect to the MSW formulae that show corrections at high neutrino energies and/or high medium density and which reduce to the latter when the scattering probability is negligible.
 In the end we present a numerical analysis for ultra high energy neutrinos propagating through the Earth and we analyze the related $CP$ and $CPT$ violations.

The paper is structured as follows: in Section~\ref{sec_interactions}, we consider  the neutrino-lepton interactions within the hypothesis that the number of neutrinos is preserved.
In Section~\ref{sec_scheme}, we discuss the general aspects of the neutrino propagation in the matter when collisions are considered and, in Section~\ref{sec_operator} we introduce the Hamiltonian and the evolution operator for neutrino-lepton scattering.
Section~\ref{sec_results}  is devoted to the analysis of the results.
First (Sub Sec.~\ref{subsec_oscillations}) we present and discuss new oscillation formulae that are valid in the region of high neutrino energy/high electron density.
Soon after, in Sub Sec.~\ref{subsec_decoherence}, we discuss the decoherence effects induced on neutrinos by the processes and, at the end, in Sub Sec.~\ref{subsec_comparison} we present some numerical prediction that could be tested in future experiments.
Section~\ref{sec_conclusion} is devoted to conclusions.

\section{Neutrino-Lepton Interactions}
\label{sec_interactions}

To begin with, let us discuss the salient assumptions of our analysis.
At first, for the sake of simplicity, we focus on the case of a neutrino with two possible flavor states $\nu_{e}, \nu_{\mu}$ propagating in matter.
This assumption is not crucial, and the approach that we use can be easily generalized to the more realistic three flavors case. Yet this would render the calculations more involved and would add no relevant physical insights to the discussion. For this reason we prefer to limit ourselves, in the present work, to the two--flavor mixing.
The second assumption is that the number of neutrinos remains unchanged in time.
This means that we are neglecting \textit{all} processes involving the emission or absorption of neutrinos, as muon decay $\mu^{-} \rightarrow e^{-} + \bar{\nu}_e + \nu_{\mu}$ so that it still makes sense to speak about the oscillations of a single neutrino.
Moreover, we consider that the interaction with the nucleons of the medium can be neglected, and retain the purely leptonic interactions with electrons $e^{-}$ and muons $\mu^{-}$ only.
Neutrino--nucleon and neutrino--nuclei interactions would indeed lead to more involved calculations, obscuring the essential physical insights that can already be grasped in our simpler setting.
In addition, as far as processes involving the emission or absorption of neutrinos can be neglected, the interaction of nucleons and nuclei with neutrinos is the same for all flavors, so that the oscillations are unaffected.
 Finally, we assume that collisions always involve a single neutrino and a single charged lepton so that $n$--body interactions with $n > 2$, whose cross-section is usually much smaller than those with $n=2$, can be neglected.

Summarizing, all the reactions we consider must then have the form
\begin{equation}\label{ParadigmReaction}
  \nu_{\alpha} + l \leftrightarrow \nu_{\beta} + l' \ \ \ \ \ \  \bar{\nu}_{\alpha} + l \leftrightarrow \bar{\nu}_{\beta} + l'
\end{equation}
where $\alpha, \beta ,l , l' = e, \mu$ and the bar denotes antineutrinos.
The form of Eq.~\eqref{ParadigmReaction} includes all the scatterings $(\nu_{\alpha}, \bar{\nu}_{\alpha}) + l \leftrightarrow (\nu_{\alpha}, \bar{\nu}_{\alpha}) + l $, and two additional reactions, one for neutrinos and one for antineutrinos. These are
\begin{equation}\label{AdditionalReaction}
  \nu_{e} + \mu^{-} \leftrightarrow \nu_{\mu} + e^{-}  \ \ \ \ \ \ \bar{\nu}_{e} + e^{-} \leftrightarrow \bar{\nu}_{\mu} + \mu^{-}
\end{equation}
and they are, apart from scatterings, the only processes of the form \eqref{ParadigmReaction} which respect lepton number conservation. In the mediums we shall be treating, the interactions of neutrinos with muons can usually be neglected, both due to the muon instability and a negligible muon density (compared to the electron density).

Depending on the neutrino energy $E_{\nu}$, and to a lesser extent on the medium density, we can distinguish two different regimes.
When low energy neutrinos are considered, a large number of constituents of the medium participates in the weak interaction, and the forward scattering amplitudes due to them add up coherently~\cite{MSW2, Lewis}.
Their collective action amounts to a (flavor--dependent) neutrino refractive index $n_{\alpha} (\pmb{x})$, which can also be understood in terms of an average potential term in the neutrino hamiltonian~\cite{GiuntiKim}
\begin{equation}\label{MatterPotential}
  V_{\alpha}(\pmb{x},t) =  \pm \delta_{\alpha e} \sqrt{2} G_F N_e (\pmb{x},t) \ .
\end{equation}
Here $\delta_{ae}$ is the Kronecker delta, $G_F$ is the Fermi constant and $N_e (\pmb{x},t) $ is the electron number density of the medium.
In Eq.~\eqref{MatterPotential} the (minus) plus sign is reserved for (anti--) neutrinos.
The coherent phenomenon described by Eq.~\eqref{MatterPotential} is the source of the MSW effect~\cite{MSW1, MSW2}.
This pure neutrino picture breaks down as the neutrino energy $E_{\nu}$ or the medium density $N_e $ are increased.
Indeed, since the total cross-section for neutrino--lepton scattering grows linearly with the energy~\cite{Vannucci, Note1.5} $\sigma_{\nu l} \propto E_{\nu}$, the number of scattering events grows, and the evolution is no longer coherent.
In most of the cases of interest, even at comparably high neutrino energies $E_{\nu} \sim 1 TeV$, the cross-section remains so small that only a few scattering events can take place during the neutrino propagation.
This can be seen immediately by evaluating the neutrino mean free path $l = (N_e \sigma)^{-1}$ for the density profiles $N_e$ of interest.
On this basis, one usually discards the second-order effects altogether, and only retains the matter potential of Eq.~\eqref{MatterPotential}.
The major drawback in doing so is that the neutrino evolution remains unitary, just as in the vacuum, so that no dissipative effects occur.
In order to encompass the dissipative effects due to neutrino--matter interactions, we shall analyze the effect of collisions on the neutrino propagation.

\section{Neutrino propagation with collisions: general scheme}
\label{sec_scheme}

In our computation we neglect all the processes implying the emission or absorption of neutrinos and therefore we consider a constant total number of neutrinos. We neglect the interaction with nucleons and we consider only the leptonic interactions with electrons and muons. We consider the system made by the neutrino and the charged lepton that take part in the scattering process as an isolated system. This leads a unitary global evolution of the two particle system, due to the scattering. We note that the total system evolves unitarily, while the neutrino evolution is dissipative.

The neutrino propagation in matter can be characterized by two evolution operators, $U_0 (t)$ and $U_W$.
The first, associated to the mixing Hamiltonian and the matter potential of Eq.~\eqref{MatterPotential}, acts on the neutrino degrees of freedom alone, and corresponds to the unitary evolution before and after the collisions.
The second describes the collision processes, and couples the neutrino  with a charged lepton of the medium.
The crucial point is that the collisions occur in a narrow spacetime region, in which the effect of $U_0 (t)$ can be neglected, so that we can schematize the neutrino propagation as a sequence of (nearly) instantaneous collisions spaced out by intervals of coherent evolution.

In principle, neutrinos may take part in an arbitrary number of collisions. Then, once that the initial neutrino state is specified $\rho_\nu(t_0)$, we define the initial state $\rho (t_0)$ of the system (neutrino + medium) as \cite{Note1.625}
\begin{equation}
\label{ancillar_state}
\rho(t_0)=\rho_\nu(t_0) \otimes \left(\bigotimes_{k=1}^{\infty}\rho_k (t_0) \right)
\end{equation}
where each $\rho_k (t_0)$ denotes the initial state of a single lepton in the medium.
 We limit our analysis to the interactions of neutrinos with leptons, and we neglect the probability for a neutrino to interact with a free muon in the medium (since the  muons are unstable and the muon density is usually negligible with respect to the electron density) with respect to the one with an electron. Therefore,   all $\rho_k (t_0)$   describe electronic states.

 Having defined the initial state, as a first step, one simply evolves the density matrix according to the evolution operator $U_0 (t) $ up to a time right before the first scattering $t_{I_1}$, to obtain $\rho (t_{I_1})$.
From this, the density matrix of the system right after the scattering $\rho (t_{F_1})$ is related to the density matrix right before the scattering $\rho(t_{I_1})$ via
\begin{equation}\label{DensityMatrixEvolution}
\rho (t_{F_1}) = U_{W_1} \rho (t_{I_1}) U_{W_1}^{\dagger} \ .
\end{equation}
where $U_{W_k}$ couple the degrees of freedom of the neutrino and the $k$-th lepton of the medium.
After the scattering event, the density matrix evolves again under the action of $U_0 (t)$.
These steps are repeated as many times as the number of collisions that occur, each time coupling the neutrino with a different charged lepton.
At each time $t$ the neutrino state can be recovered  by tracing out the charged lepton degrees of freedom $\rho_{\nu} (t) = Tr_l (\rho (t))$.
It is worth to note that, while the evolution of the complete density matrix (neutrino plus charged leptonic  states) is unitary, the evolution of the reduced density matrix associated to the sole neutrino is not.
This implies that differentiating  $\rho_{\nu} (t)$ with respect to time, one obtains, along with the Hamiltonian, further dissipative terms that induce decoherence effects.

At the end of the process, the neutrino density matrix can generally be written as
\begin{equation}\label{NeutrinoDensityMatrix}
  \rho_{\nu}(t) = \sum_{n = 0}^{\infty} P_{n} (t) \rho_{\nu}^{(n)} (t) \ ,
\end{equation}
where $P_{n}(t)$ is the probability that after a time $t$, $n$ scattering events, involving the same neutrino, have taken place, and  $\rho_{\nu}^{(n)} (t)$ is the associated density matrix.
If the medium is homogenous, the probability of having a scattering, at any time, is independent of $t$ and $n$, so that $P_n (t)$ is a Poisson distribution
$P_n (t) = \frac{1}{n!}(\Gamma t)^n e^{- \Gamma t}$~\cite{Note1.75}.
Since the instants $\{t_1, ..., t_n \}, 0 \leq t_i \leq t$ at which the collisions occur are randomly distributed in the interval $[0,t]$, the density matrix $\rho_{\nu}^{(n)} (t)$ for $n$ collisions involves a statistical average
\begin{equation}\label{NCollisionMatrix}
  \rho_{\nu}^{(n)} (t) =\!\! \int_{t_1 = 0}^{t}\!\int_{t_n > t_{n-1} ... > t_2 > t_1}^{t} \! \! \! \! \! \! \! \! \! \! \! \! \! \! \! \! \! \! \! \! \! \! \! \! \! \! \! \! \! \! \! \! \! \! \! \! dt_1 ... dt_n  R_n(t_1,...,t_n) \eta_{\nu} (t; t_1,...,t_n) .\!\!
\end{equation}
In equation \eqref{NCollisionMatrix},  $\eta_{\nu} (t; t_1,...,t_n)$ is the neutrino density matrix at time $t$, after $n$ collisions occurring exactly at the times $\{t_1, ..., t_n \}$  while  $R_n(t_1,...,t_n)$ stands for the joint probability distribution of the $n$ collisions occurring at the times  $\{t_1, ..., t_n \}$, with normalization
\begin{equation}\label{Normalization1}
  \int_{t_1 = 0}^{t}\int_{t_n > t_{n-1} ... > t_2 > t_1}^{t} dt_1 ... dt_n R_n(t_1,...,t_n) = 1
\end{equation}
corresponding to the certainty that $n$ collisions occur in the interval $[0,t]$.

As it can be seen from the above discussion, the determination of the neutrino density matrix with growing $n$ becomes soon intractable analytically, and, in the general case, for a large number of collisions some approximation must be invoked.
Nevertheless, for a very large set of physical situations of interest the probability of having $2$ or more collisions is extremely small and can be neglected.
To give an idea, for electronic densities of the order of $N_e = N_A cm^{-3}$, with $N_A$ the Avogadro's number, and neutrino energies $E \sim 1$ GeV, the interaction rate $\Gamma = N_e \sigma_T$, where $\sigma_T$ is the total cross section, is as low as $\Gamma \sim  10^{-7} s^{-1}$.
In this case eq.~\eqref{NeutrinoDensityMatrix} simplifies to
\begin{equation}\label{NeutrinoDensityMatrixSimpl}
  \rho_{\nu} (t) = (1 - P_1(t)) \rho_{\nu}^{(0)} (t) + P_1(t) \rho_{\nu}^{(1)} (t) \
\end{equation}

\section{Neutrino--Lepton Scattering: Hamiltonian and Evolution operator}
\label{sec_operator}

In this section we derive the evolution operator describing the scattering. We first establish the general form of the scattering Hamiltonian on the basis of the possible reactions. We compare the scattering matrix obtained via quantum field theoretical calculations with the scattering matrix that would be produced by a quantum mechanical interaction Hamiltonian. Computing the scattering amplitudes allows us to derive the components of the scattering Hamiltonian. For the computation of the evolution operator, we observe that it has the same symmetries as the Hamiltonian, and then use both its expression in terms of the Hamiltonian and its probabilistic interpretation in order to find its precise form. In the following we consider only the forward scattering of the neutrinos, and we assume that the $4$-momentum transfer, in all the reactions considered, is negligible with respect to the mass of the weak bosons $m_W, m_Z$.

Due to the assumptions made, both the neutrino and the charged lepton taking part in the reactions of Eq.~\eqref{ParadigmReaction} can be thought of as two--level systems.
The neutrino (antineutrino) can be either in the electronic $\ket{\nu_e}$ ($\ket{\bar{\nu}_e}$) or in the muonic $\ket{\nu_{\mu}}$ ($\ket{\bar{\nu}_\mu}$) state, while, similarly, we can regard the electron $\ket{e}$ and the muon $\ket{\mu}$ respectively as the ground and excited state of the charged lepton $\ket{l}$.
Since all the processes of Eq.~\eqref{ParadigmReaction} involve a neutrino (or an antineutrino) and a charged lepton in both the initial and final state, a useful basis in which the processes can be described is given by
\begin{equation}\label{InteractionStates}
  \ket{\nu_l, l'} \doteq \ket{\nu_l} \otimes \ket{l'} \ ,
\end{equation}
with $l,l' = e , \mu$. The lepton states $\ket{l'}$ are obviously referred to the particular charged lepton of the medium that takes part in the interaction.
A couple of comments are in order.
For our purposes the exact spacetime dependence of the states of Eq.~\eqref{InteractionStates} is irrelevant, as long as they are sharply peaked around a given 4--momentum.
The essential requirement is that they contain all the kinematical information needed to characterize the reactions of Eq.~\eqref{ParadigmReaction}. Within our assumptions, this amounts to specifying the neutrino energy $E_{\nu}$ and the charged lepton spin projection (as the neutrino helicity is fixed \cite{Note2}). The generic state is then written
\begin{equation}\label{InteractionStatesPrime}
 \ket{\nu_l,l'}^{(h)}_{E_{\nu}}    \  ,
\end{equation}
where $E_{\nu}$ is the neutrino energy and $h=\pm 1$ is the charged lepton spin projection.
In principle one might choose any reference frame to study the reactions, and then the states of Eq.~\eqref{InteractionStatesPrime} would be characterized by the neutrino and the charged lepton 4--momenta $p_{(\nu)}^{\mu}, p_{(l)}^{\mu}$ in this frame.
One should then enforce the total 4--momentum conservation for each of the processes in  Eq.~\eqref{ParadigmReaction} by suitably defining the inner products between the states.
  However, it is much more convenient to work in the rest frame of the medium, which simplifies extremely both the kinematics and the notation.
Denoting by $p_{\alpha}^{\mu}, p_{l}^{\mu}$ (with $\alpha = \nu_e,\nu_{\mu}$ and $l=e,\mu$) respectively the 4-momentum of the incoming neutrino and of the target charged lepton, we have that the latter is at rest in the chosen frame, so that $p_{l}^{\mu} \equiv (m_l,0,0,0)$, with $m_l$ its mass. The assumption that the target charged lepton is at rest in the medium rest frame requires that the temperature of the medium be negligible with respect to the neutrino energy.

Choosing the third spatial axis to coincide with the direction of motion of the incoming neutrino, its momentum reads $p_{\alpha}^{\mu} \equiv (E_{\alpha},0,0,p_{\alpha})$, with $p_{\alpha} \simeq E_{\alpha}$. Among the reactions of Eq. \eqref{ParadigmReaction} we can distinguish elastic scattering, where the initial and the final charged lepton are the same $l=l'$, and quasi--elastic scattering, where the charged lepton is converted to another kind $l\neq l'$.
For the elastic scattering one can enforce the hypothesis of vanishing $4$-momentum transfer, so that the neutrino preserves both its initial energy and its initial direction of motion. This hypothesis is in line with the derivation of the MSW effect, and corresponds to the requirement that the medium remains unaltered by the interaction with neutrinos.
The condition is $q^{\mu} = p_{\beta}^{\mu} - p_{\alpha}^{\mu} \simeq 0$, where $p_{\beta}^{\mu}$ is the 4-momentum of the outgoing neutrino and $p_{\alpha}^{\mu}$ is the 4-momentum of the incoming neutrino.
This means that the outgoing neutrino is emitted within a small solid angle $\delta \Omega$ around the direction of the incoming neutrino.
This also implies a vanishing change in the neutrino energy, so that the 4-momentum of the outgoing neutrino is
\begin{equation}\label{Final4momentum}
  p_{\beta}^{\mu} \simeq p_{\alpha}^{\mu} \equiv (E_{\alpha},0,0,p_{\alpha})  \ .
\end{equation}
In addition, since the $3$-momentum is retained by the neutrino $\pmb{p}_{\alpha} = \pmb{p}_{\beta}$, by conservation of the total $3$-momentum, the spatial momentum of the outgoing charged lepton is zero $\pmb{p}_{l'} \simeq 0$.
This in turn implies that the $4$-momentum of the outgoing charged lepton is
\begin{equation}\label{Final4momentum2}
  p^{\mu}_{l'} \simeq (m_{l'},0,0,0) \ .
\end{equation}
It is obvious that within these assumptions the neutrino energy and the charged lepton spin projection suffice to determine the kinematics of the reaction completely.

On the other hand, for the quasi--elastic reactions it is not kinematically possible to have both a vanishing $4$-momentum transfer and forward scattering. The only requirement we can impose is that the neutrino is forward scattered, i.e. that it preserves its original direction of motion. The kinematics in the rest frame of the medium are worked out in detail in the appendix (\ref{FirstAppendix}). The crucial point is that the incoming neutrino necessarily transfers (or receives) a part of its energy to (from) the medium $E_{\alpha} \neq E_{\beta}$. The states of equation \eqref{InteractionStatesPrime} are still sufficient to describe the reaction, but we need at least two sets $\ket{\nu_l,l'}^{(h)}_{E_{\alpha}}, \ket{\nu_l,l'}^{(h)}_{E_{\beta}}$ corresponding to distinct neutrino energies.

If one relaxes the assumptions of vanishing 4-momentum transfer and small scattering angle, the parameters $E_{\nu}, h$ no longer suffice to specify the states of Eq.~\eqref{InteractionStates}, and the complete kinematics of the reaction must be encoded in the states.
We defer these complications to later studies, and, from now on, we shall always work in the 8--dimensional Hilbert space $\mathcal{H}^{(h)}_{E_{\alpha}} \oplus \mathcal{H}^{(h)}_{E_{\beta}}$ spanned by the vectors of Eq.~\eqref{InteractionStatesPrime} corresponding to the initial neutrino energy $E_{(\alpha)}$ in the rest frame of the medium and the charged lepton spin projection $h$. The final neutrino energy is uniquely determined both for the elastic reactions ($E_{\beta} = E_{\alpha}$) and the quasi-elastic reactions ($E_{\beta} \neq E_{\alpha}$) (see the appendix (\ref{FirstAppendix})).
The situation for the anti--neutrinos, whose states we denote $\ket{\bar{\nu}_l,l'}^{(h)}_{E_{\alpha}}$, is analogous.

Having established which states are relevant for our analysis, we can now determine the explicit forms of $U_0$ and $U_{W_k}$ on the basis states of Eq. \eqref{InteractionStatesPrime}.
The operator $U_0$ determines the evolution of the neutrino states between two successive collisions and can be put in the following form
\begin{equation}\label{CoherentOperator}
  U_0 (t) = e^{- \imath (H_M + H_{MSW}) t} \ .
\end{equation}
The first Hamiltonian term is that of mixing, which is responsible for the neutrino oscillations.
Dropping terms proportional to the identity, it can be written as~\cite{CapolupoCPTdiss}
\begin{eqnarray}\label{MixingHamiltonian}
  H_M = -\omega_0 \left(\cos 2 \theta \sigma_{\nu}^{z} - \sin 2 \theta \sigma_{\nu}^{x} \right) \otimes \mathbb{1}_l
\end{eqnarray}
where $\omega_0 = \frac{\Delta m^2}{4 E_{\nu}}$, $\theta$ is the two--flavor mixing angle and $\Delta m^2 = m_2^2 - m_1^2$ is the squared neutrino mass difference.
The subscripts $\nu$ and $l$ refer to the neutrino and the lepton subspaces, so that $\sigma_{\nu}^{j}$ ($\sigma_{l}^{j}$) denotes the $j$-th Pauli operator on the neutrino (leptonic) subspace.
The Hamiltonian \eqref{MixingHamiltonian} obviously depends on the neutrino energy $E_{\nu}$ via $\omega_0$, so that we have two copies  $H_{M,E_{\alpha}}, H_{M,E_{\beta}}$ of $H_M$, one for each energy subspace.
The Hamiltonian in Eq.~\eqref{MixingHamiltonian}, assuming no CP violating phase, is valid for both neutrinos and antineutrinos.
The second term is the MSW matter potential of Eq.~\eqref{MatterPotential}, which can be written as
\begin{equation}\label{MSWHamiltonian}
  H_{MSW} = \pm \frac{\sqrt{2}}{2} G_F N_e  \sigma_{\nu}^z \otimes \mathbb{1}_l \ .
\end{equation}
The factor $\frac{1}{2}$ in Eq.~\eqref{MSWHamiltonian} comes from the subtraction of a term proportional to the identity, namely $\pm \frac{\sqrt{2}}{2} G_F N_e  \mathbb{1}_{\nu} \otimes \mathbb{1}_l $ (compare with equation 9.14 in Ref.~\cite{GiuntiKim}) and the (minus) plus has to be chosen for (anti--)neutrinos.
Since neither the mixing Hamiltonian $H_M$, nor the the matter potential $H_{MSW}$ change the neutrino energy, the $U_0$ operator can be written in the $8$-dimensional basis as
\begin{equation}\label{MSWEvolution8}
  \mathcal{U}_{0}(t) =  \left( \begin{array}{cccc}
 U_{0 \alpha} (t)& 0  \\
  0 & U_{0 \beta} (t)
 \end{array} \right) \ ,
\end{equation}
where $U_{0 \alpha} (t),  U_{0 \beta} (t)$ are the $4 \times 4$ matrices given by Eq.~\eqref{CoherentOperator} corresponding respectively to the neutrino energies $E_{\alpha}, E_{\beta}$.

On the other hand, concerning the collisions, the operators $U_{W_k}$ can be written as
\begin{equation}
\label{UW}
U_{W_k}=e^{-\imath H_W \tau}
\end{equation}
where $\tau$ is the time scale of the scattering process and $H_W$ is the Hamiltonian governing the collision. Notice  that Eq. \eqref{UW} that describes contribution of the scattering processes (quadratic
in the interaction Hamiltonian) has the form of a unitary evolution operator. As
a consequence, the time-derivative of the density matrix contains a commutator
of $H_W$ and the density matrix, instead of the standard anti-commutator, that
appears in the kinetic approach to neutrino oscillations. We point out that this seeming contradiction arises because Eq. \eqref{UW} refers to the evolution of the complete density matrix of the neutrino plus charged lepton state and not to the neutrino alone. This means that Eq. \eqref{UW} describes the evolution of both the physical system (the neutrino) and the reservoir (the charged lepton). For this reason, at this level, the evolution of the total state is unitary. Moreover, equation \eqref{UW} is meant to describe the single scattering event, meaning that no information on the collision rate, related to the cross section, is contained within it. The collision rate, which of course, depends quadratically on the interaction Hamiltonian, is instead encoded in the probabilities $ P_{n} (t)$ of Eq. \eqref{NeutrinoDensityMatrix}. On the contrary, in the kinematic approach to neutrino oscillations, the evolution of the reduced density matrix, associated to neutrinos alone, is considered. This implies a non--unitary evolution which takes the form described in the refs. \cite{Vlasenko2013,Blaschke2016,Raffelt2018} and the presence of anti--commutators in place of the commutator deriving from Eq. \eqref{UW}. We remark that the comparison between our approach and the ones presented in \cite{Vlasenko2013,Blaschke2016,Raffelt2018} must be done at the level of the \emph{reduced} density matrix and not for the evolution of the complete density matrix which is instead necessarily unitary. In terms of the reduced density matrix associated only to the neutrino, the two approaches are compatible.
Taking into account all the possible processes involving a neutrino and a charged leptonic state in Eqs.~\eqref{ParadigmReaction} and \eqref{AdditionalReaction}, $H_W$ can be written as

\begin{widetext}
 \begin{equation*}\label{WHamiltonian81}
  H_{W}^{(h)} =  \left( \begin{array}{cccccccc}
 \alpha_1^{(h)} & 0 & 0 & 0 & 0 & 0 & 0 & 0  \\
  0 & \alpha_2^{(h)} & 0 & 0 & 0 & 0 & 0 & 0 \\
  0 & 0 & \alpha_3^{(h)} & 0 & 0 & \beta^{(h)} & 0 & 0 \\
  0 & 0 & 0 & \alpha_4^{(h)} & 0 & 0 & 0 & 0 \\
  0 &0 & 0 & 0 & \alpha_5^{(h)} & 0 & 0 & 0 \\
  0 & 0 &  \beta^{(h)*} & 0 & 0 & \alpha_6^{(h)} & 0 & 0 \\
  0 &0 & 0 & 0 & 0 & 0 & \alpha_7^{(h)} & 0 \\
  0 &0 & 0 & 0 & 0 & 0 & 0 & \alpha_8^{(h)}
 \end{array} \right)\,, \ \ \ \ \ \   \bar{H}_{W}^{(h)} =  \left( \begin{array}{cccccccc}
 \bar{\alpha}_1^{(h)} & 0 & 0 & 0 & 0 & 0 & 0 & \gamma^{(h)} \\
  0 & \bar{\alpha}_2^{(h)} & 0 & 0 & 0 & 0 & 0 & 0 \\
  0 & 0 & \bar{\alpha}_3^{(h)} & 0 & 0 & 0 & 0 & 0 \\
  0 & 0 & 0 & \bar{\alpha}_4^{(h)} & 0 & 0 & 0 & 0 \\
  0 &0 & 0 & 0 & \bar{\alpha}_5^{(h)} & 0 & 0 & 0 \\
  0 & 0 &  0 & 0 & 0 & \bar{\alpha}_6^{(h)} & 0 & 0 \\
  0 &0 & 0 & 0 & 0 & 0 & \bar{\alpha}_7^{(h)} & 0 \\
  \gamma^{(h)*} &0 & 0 & 0 & 0 & 0 & 0 & \bar{\alpha}_8^{(h)}
  \end{array} \right) \ .
\end{equation*}
\end{widetext}
Here $\bar{H}_W$ is the Hamiltonian for the antineutrinos, the quantities $\alpha, \bar{\alpha}, \beta, \gamma$ are parameters to be determined and the apex $h$ refers to the charged lepton spin projection. The asterisk denotes complex conjugation. The following convention is adopted
\begin{eqnarray*}
 && \ket{\nu_e,e}^{(h)}_{E_{\alpha}} \rightarrow \pmb{e}_1 \,,\ \ \ \ \ \ \ket{\nu_{e},\mu}^{(h)}_{E_{\alpha}} \rightarrow \pmb{e}_2 \\
 &&\ket{\nu_{\mu},e}^{(h)}_{E_{\alpha}} \rightarrow \pmb{e}_3 \,, \ \ \ \ \ \
 \ket{\nu_{\mu},\mu}^{(h)}_{E_{\alpha}} \rightarrow \pmb{e}_4 \\
 && \ket{\nu_e,e}^{(h)}_{E_{\beta}} \rightarrow \pmb{e}_5 \,,\ \ \ \  \ \ \ket{\nu_{e},\mu}^{(h)}_{E_{\beta}} \rightarrow \pmb{e}_6 \\
 &&\ket{\nu_{\mu},e}^{(h)}_{E_{\beta}} \rightarrow \pmb{e}_7 \,, \ \ \ \ \ \
 \ket{\nu_{\mu},\mu}^{(h)}_{E_{\beta}} \rightarrow \pmb{e}_8
\end{eqnarray*}
where $\pmb{e}_j$ is the basis vector with a $1$ in the $j$-th position and $0$ elsewhere, and similarly for the antineutrinos.

In order to fix these quantities, we need to analyze the field theoretical amplitude for the reactions of Eq.~\eqref{ParadigmReaction}.
The Hamiltonian and the related evolution operators of Eq. \eqref{UW} are intended with respect to the basis states of Eq. \eqref{InteractionStatesPrime} formed by the neutrino and the specific charged lepton (the $k-th$) taking part in the collision. Assuming that the field theoretical 2-particle states $\ket{p_1,p_2}$ are normalized as $\braket{p_1,p_2|p'_1,p'_2} = (2\pi)^3 \delta^3 (\pmb{p'_1}- \pmb{p_1})  (2\pi)^3(\pmb{p'_2}- \pmb{p_2}) $ , the scattering amplitude reads~\cite{PDG}
{\small
\begin{equation}\label{Scattering Amplitude}
  \bra{p'_1,p'_2}\! S\! \ket{p_1,p_2}\! =\! I\! -\! \imath  \delta^4 (p_1 + p_2 - p'_1 - p'_2) \frac{(2\pi)^4 M(p,p')}{\sqrt{16 E_1 E_2 E'_1 E'_2}}
\end{equation}}
where $I$ is the identity and $M(p,p')$ is the Lorentz-invariant amplitude for the process.
The matrix elements of the interaction Hamiltonian can be deduced by comparing Eq.~\eqref{Scattering Amplitude} with the scattering matrix that would be produced by an interaction potential (see for instance \cite{Shapiro}), which to first order reads
\begin{equation}\label{QMScatteringAmplitude}
  \bra{f} S \ket{i} = I - 2 \pi i \delta(E_f - E_i) \bra{f} H_I \ket{i} \ .
\end{equation}
Here $\ket{i}, \ket{f}$ denote the initial and final states, $E_i,E_f$ the total initial and final energy and $H_I$ is the interaction Hamiltonian.
Comparing Eqs. \eqref{Scattering Amplitude} and \eqref{QMScatteringAmplitude} we find
\begin{equation}\label{IntermediateStep1}
\bra{f}\! H_I\! \ket{i}\! =\! (2 \pi)^3 \!\delta^3 (\pmb{p'_1} + \pmb{p'_2} - \pmb{p_1} - \pmb{p_2}) \frac{{M(p,p')}}{\sqrt{16 E_1 E_2 E'_1 E'_2}} .
\end{equation}
The Eq.~\eqref{IntermediateStep1} yields the matrix elements of the interaction Hamiltonian on the field theoretical states $\ket{p_1,p_2}$.
To go further we need the matrix elements of the interaction Hamiltonian on the states of Eq.~\eqref{InteractionStatesPrime} which are normalized to unity.
This is fundamental if one wishes to obtain the transition probability to a definite final state and thus derive a well--defined quantum mechanical operator $H_I$.
To renormalize the two particle states, we follow Ref.~\cite{Weinberg} and enclose the system in a box with large volume $V$.
As we only consider states which satisfy 4-momentum conservation, as the states of Eq.~\eqref{InteractionStatesPrime} automatically do, the Dirac delta in \eqref{IntermediateStep1} is just $\delta^3 (0) = \frac{V}{2 \pi^3}$.
In addition, the singular normalization of the 2-particle states becomes
\begin{equation}\label{RegularNormalization}
 \braket{p_1,p_2|p'_1,p'_2} = V^2 \delta_{\pmb{p_1}, \pmb{p'_1}} \delta_{\pmb{p_2}, \pmb{p'_2}}
\end{equation}
where now $\delta_{\pmb{p}, \pmb{q}}$ is a Kronecker delta.
The two particle states are then normalized to unity if all of them are multiplied by $\frac{1}{V}$.
To find the normalized matrix elements we just divide Eq.\eqref{IntermediateStep1} by $V^2$, so to obtain
\begin{equation}\label{NormalizedHamiltonian}
  \bra{f} H_I^{(norm)} \ket{i} = \frac{{M(p,p')}}{V \sqrt{16 E_1 E_2 E'_1 E'_2}} \ .
\end{equation}
The Lorentz-invariant amplitude $M(p,p')$ for the reactions of Eq.~\eqref{ParadigmReaction} can be computed employing the Feynman rules for the electroweak interaction. The calculation of the amplitudes is performed in the appendix (\ref{FirstAppendix}).

We show the derivation of $\alpha_1^{(+)}$ explicitly, but all the parameters of $H_W$ and $\bar{H}_W$ can be computed in a similar fashion. To find $\alpha_1^{(+)}$ we take $\ket{f} \equiv \ket{i} = \ket{\nu_e,e}^{(+)}_{E_{\nu}}$ where we have taken into account that this reaction does not change the neutrino energy $E_{\alpha} = E_{\beta} = E_{\nu}$. Inserting the amplitude from Eq.\eqref{FirstAmplitude} in Eq. \eqref{NormalizedHamiltonian}, we deduce
\begin{eqnarray*}
 \alpha_1^{(+)} &=& \frac{-8\sqrt{2}G_F E_{\nu} m_e \left(\sin^2 \theta_W + \frac{1}{2}\right)}{4V \sqrt{E_{\nu}^2 m_e^2}} \\ &=& - \frac{2\sqrt{2} G_F}{V} \left( \sin^2 \theta_w + \frac{1}{2} \right) \  .
\end{eqnarray*}
By the same steps we can fix all the parameters appearing in the scattering Hamiltonian. We quote the result for the reader's convenience.
\begin{eqnarray}\label{HamiltonianElements}
 \nonumber \alpha_1^{(+)} &=& \alpha_4^{(+)} = \alpha_5^{(+)} = \alpha_8^{(+)} = - \frac{2\sqrt{2} G_F}{V} \left( \sin^2 \theta_W + \frac{1}{2} \right) \\
  \nonumber \alpha_2^{(+)} &=& \alpha_2^{(+)} = \alpha_6^{(+)} = \alpha_7^{(+)} = - \frac{2\sqrt{2} G_F}{V} \left( \sin^2 \theta_W - \frac{1}{2} \right) \\
  \nonumber \beta^{(+)} &=& \frac{4 G_F}{V} \frac{m_e}{\sqrt{m_e^2 + m_{\mu}^2}} \\
   \nonumber \alpha_j^{(-)} &=&   - \frac{2\sqrt{2} G_F}{V}  \sin^2 \theta_W   \ \ \ \ \forall \ j=1,...,8  \\
\nonumber  \beta^{(-)} &=& 0 \\
\nonumber \bar{\alpha}_1^{(+)} &=& \bar{\alpha}_4^{(+)} = \bar{\alpha}_5^{(+)} = \bar{\alpha}_8^{(+)} = \frac{2\sqrt{2} G_F}{V} \left(\frac{3}{2} - \sin^2 \theta_W \right) \\
\nonumber \bar{\alpha}_2^{(+)} &=& \bar{\alpha}_3^{(+)} = \bar{\alpha}_6^{(+)} = \bar{\alpha}_7^{(+)} = \frac{2\sqrt{2} G_F}{V} \\
\nonumber \gamma^{(+)} &=& \frac{4 G_F}{V} \frac{m_e}{\sqrt{m_e^2 + m_{\mu}^2}} \\
\nonumber \bar{\alpha}_1^{(-)} &=& \bar{\alpha}_4^{(-)} = \bar{\alpha}_5^{(-)} = \bar{\alpha}_8^{(-)} = -\frac{2\sqrt{2} G_F}{V} \sin^2 \theta_W \\
\nonumber \bar{\alpha}_2^{(+)} &=& \bar{\alpha}_3^{(+)} = \bar{\alpha}_6^{(+)} = \bar{\alpha}_7^{(+)} = 0 \\
\gamma^{(-)} &=& 0   \ .
\end{eqnarray}

The scattering Hamiltonian shall be expedient in the determination of the corresponding evolution operator.
Notice that, the former still contains the undetermined, and in principle arbitrary, normalization volume $V$ while the latter is still characterized by the unknown time $\tau$.
To get rid of the interacting volume $V$ and time $\tau$ it is convenient to work directly with the time evolution operator.

Naturally, $U_W$ is a unitary $8 \times 8$ matrix on the basis states of Eq.~\eqref{InteractionStatesPrime}.
Exactly as it happens for the Hamiltonians, because of the different weak processes occurring we have two distinct operators, one for neutrinos $U_W$  and one for antineutrinos $\bar{U}_W$.
From Eq.~\eqref{HamiltonianElements}, it is clear that the only non--zero elements are those on the main diagonal $U_{W, jj}, \bar{U}_{W, jj}$ with $j=1,...,8$ and the elements $U_{W, 36}, U_{W, 63}, \bar{U}_{W, 18}, \bar{U}_{W, 81}$. In order to fix the evolution operator, we start by splitting the scattering Hamiltonian conveniently. Defining $\alpha_{\pm}^{(+)} = \frac{\alpha_{1}^{(+)}\pm \alpha_2^{(+)}}{2}$, we can write

\begin{widetext}
 \begin{equation}\label{SimplHamiltonian1}
  H_W = \alpha_{+}^{(+)} \mathbb{1} + K^{(+)}_{W}
 \,, \ \ \ \ \ \ \ \ \ \ \ \
  K^{(+)}_{W} =  \left( \begin{array}{cccccccc}
 \alpha_{-}^{(+)} & 0 & 0 & 0 & 0 & 0 & 0 & 0  \\
  0 & \alpha_{-}^{(+)} & 0 & 0 & 0 & 0 & 0 & 0 \\
  0 & 0 & \alpha_{-}^{(+)} & 0 & 0 & \beta^{(+)}& 0 & 0 \\
  0 & 0 & 0 & \alpha_{-}^{(+)} & 0 & 0 & 0 & 0 \\
  0 &0 & 0 & 0 & \alpha_{-}^{(+)} & 0 & 0 & 0 \\
  0 & 0 &  \beta^{(+)} & 0 & 0 & \alpha_{-}^{(+)} & 0 & 0 \\
  0 &0 & 0 & 0 & 0 & 0 & \alpha_{-}^{(+)} & 0 \\
  0 &0 & 0 & 0 & 0 & 0 & 0 & \alpha_{-}^{(+)}
 \end{array} \right) \ ,
\end{equation}

By exponentiating $U_W = e^{-i H_W \tau}$, it follows that $U_W^{(+)} = e^{-i \alpha^{(+)}_{+}\tau} V^{(+)}_{W}$, where

 \begin{equation}\label{SimplEvolution81}
  V^{(+)}_{W} =  \left( \begin{array}{cccccccc}
 e^{-i \phi} & 0 & 0 & 0 & 0 & 0 & 0 & 0  \\
  0 & e^{-i \phi} & 0 & 0 & 0 & 0 & 0 & 0 \\
  0 & 0 & e^{i \phi} \cos \beta^{(+)} \tau & 0 & 0 & -i e^{i \phi} \sin \beta^{(+)} \tau& 0 & 0 \\
  0 & 0 & 0 & e^{- i \phi} & 0 & 0 & 0 & 0 \\
  0 &0 & 0 & 0 & e^{- i \phi} & 0 & 0 & 0 \\
  0 & 0 &  -i e^{i \phi} \sin \beta^{(+)} \tau & 0 & 0 & e^{i \phi} \cos \beta^{(+)} \tau & 0 & 0 \\
  0 &0 & 0 & 0 & 0 & 0 & e^{- i \phi} & 0 \\
  0 &0 & 0 & 0 & 0 & 0 & 0 & e^{- i \phi}
 \end{array} \right) \ ,
\end{equation}
and $\phi = \alpha_{-}^{(+)}\tau$. The global phase factor $e^{-i \alpha^{(+)}_{+}\tau}$ does not affect any observable of interest and then it can be discarded.
Evidently $\sin^2 \beta^{(+)}\tau$ is the probability, once the scattering occurs, that given the initial state $\ket{\nu_{\mu},e}^{(+)}_{E_{\alpha}}$, the final state is $\ket{\nu_{e},\mu}^{(+)}_{E_{\beta}}$.
According to Fermi's golden rule, the transition probability from state $\ket{i}$ to state ${f}$ is proportional to the squared modulus of the matrix element $|\bra{f} H_I \ket{i}|^2$ of the interaction Hamiltonian. Then the relative transition probability from the initial state $\ket{\nu_{\mu},e}^{(+)}_{E_{\alpha}}$ to the final state $\ket{\nu_{e},\mu}^{(+)}_{E_{\beta}}$ is given by
\begin{equation}
 \sin^2 \beta^{(+)} \tau = \frac{|H_{W,36}^{(+)}|^2}{|H_{W,33}^{(+)}|^2 + |H_{W,36}^{(+)}|^2} = \frac{|\beta^{(+)}|^2}{|\alpha_2^{(+)}|^2 +|\beta^{(+)}|^2}  \ .
\end{equation}
Substituting the values derived in Eq.\eqref{HamiltonianElements}, we obtain
\begin{eqnarray}\label{TauEquation0}
\nonumber \sin \beta^{(+)}\tau &=& \sqrt{\frac{2m_e^2}{(m_e^2 + m_{\mu}^2)\left[\left(\sin^4 \theta_W - \sin^2 \theta_W + \frac{1}{4}\right) + \frac{2m_e^2}{m_e^2 + m_{\mu}^2} \right]}} \ \  
\end{eqnarray}
and then
\begin{eqnarray}\label{TauEquation}
\beta^{(+)} \tau &=& \arcsin \left( \sqrt{\frac{2m_e^2}{(m_e^2 + m_{\mu}^2)\left(\sin^4 \theta_W - \sin^2 \theta_W + \frac{1}{4}\right) + 2m_e^2 }}\right)  \ .
\end{eqnarray}

The Eq.\eqref{TauEquation} also allows for an immediate determination of the phase factor $\phi$. Indeed
\begin{eqnarray}
 \phi = \alpha_{-}^{(+)}\tau = \frac{\alpha_{-}^{(+)}}{\beta^{(+)}} \beta^{(+)}\tau = - \frac{\sqrt{2}}{4} \sqrt{1 + \frac{m_{\mu}^2}{m_e^2}} \arcsin \left( \sqrt{\frac{2m_e^2}{(m_e^2 + m_{\mu}^2)\left(\sin^4 \theta_W - \sin^2 \theta_W + \frac{1}{4}\right) + 2m_e^2 }}\right)\,.
\end{eqnarray}
The same analysis carried over for $U^{(+)}_W$ can also be performed for the opposite charged lepton spin projection and for the antineutrino. In particular, given the matrix elements of Eq.\eqref{HamiltonianElements}, $U^{(-)}_W$ is just a phase factor times the identity matrix. For $\bar{U}^{(+)}_W$ one arrives at the form $\bar{U}^{(+)}_W = e^{- i \bar{\alpha}^{(+)}_{+}\tau} \bar{V}^{(+)}_W$ with

 \begin{equation}\label{SimplEvolution82}
  \bar{V}^{(+)}_{W} =  \left( \begin{array}{cccccccc}
 e^{-i \bar{\phi}} \cos \gamma^{(+)}\tau & 0 & 0 & 0 & 0 & 0 & 0 &  -i e^{i \bar{\phi}} \sin \gamma^{(+)} \tau \\
  0 & e^{-i \bar{\phi}} & 0 & 0 & 0 & 0 & 0 & 0 \\
  0 & 0 & e^{-i \bar{\phi}} & 0 & 0 & 0 & 0 & 0 \\
  0 & 0 & 0 & e^{-i \bar{\phi}} & 0 & 0 & 0 & 0 \\
  0 &0 & 0 & 0 & e^{-i \bar{\phi}} & 0 & 0 & 0 \\
  0 & 0 &  0 & 0 & 0 & e^{-i \bar{\phi}} & 0 & 0 \\
  0 &0 & 0 & 0 & 0 & 0 & e^{-i \bar{\phi}} & 0 \\
  -i e^{i \bar{\phi}} \sin \gamma^{(+)} \tau &0 & 0 & 0 & 0 & 0 & 0 & e^{-i \bar{\phi}}\cos \gamma^{(+)}\tau
 \end{array} \right) \ ,
\end{equation}
where $\bar{\phi}= \bar{\alpha}_{-}^{(+)}$. The relative transition probability is fixed, also in this case, by inspecting the ratio of the matrix elements of the Hamiltonian:
\begin{equation}
 \gamma^{(+)} \tau = \arcsin \left( \sqrt{\frac{2m_e^2}{(m_e^2 + m_{\mu}^2)\left(\sin^4 \theta_W - 3\sin^2 \theta_W + \frac{9}{4}\right) + 2m_e^2 }}\right)  \ .
\end{equation}
 \end{widetext}

\section{Results}
\label{sec_results}

The analysis carried over in the previous sections provides the tools necessary to determine the evolution of the neutrino state for the propagation in a dense medium, under the assumptions of a sufficiently high energy $E_{\nu} > 1 GeV$ and purely leptonic interactions.
For an arbitrary medium, with a non--constant density profile, the evaluation of the density matrix and the related quantities of interest is an extremely complicated numerical task.
However, when a constant density profile is assumed, some predictions can be provided analytically.

\subsection{Oscillations for high electron density and/or high neutrino energy case}
\label{subsec_oscillations}

Accordingly with the MSW theory, the mixing angle in matter $\theta_m$ predicted from the effective potential of Eq.\eqref{MatterPotential} satisfies the relation
\begin{equation}\label{MatterMixingAngle}
  \sin 2 \theta_m = \frac{\omega_0 \sin 2 \theta}{\sqrt{(\omega_0 \cos 2 \theta \pm \frac{G_F N_e}{\sqrt{2}})^2 + \omega_0^2 \sin^2 2 \theta}} \ ,
\end{equation}
where $\theta$ is the mixing angle in vacuum.
When $G_F N_e \gg \omega_0$ the equation above becomes, to the leading order
\begin{equation}\label{SmallMatterMixingAngle}
  \sin 2 \theta_m \simeq \epsilon \sin 2 \theta
\end{equation}
where $\epsilon = \frac{\sqrt{2}\omega_0}{G_F N_e}$.
Then, at high density and/or high neutrino energy, the matter potential inhibits the oscillations, whose probability is proportional to $\sin 2 \theta_m$, with the mixing angle falling down with growing energy and density $(\theta_m \propto \frac{1}{N_e E})$.

\begin{figure}\label{OscPlot}
\centering
  \includegraphics[width=\linewidth]{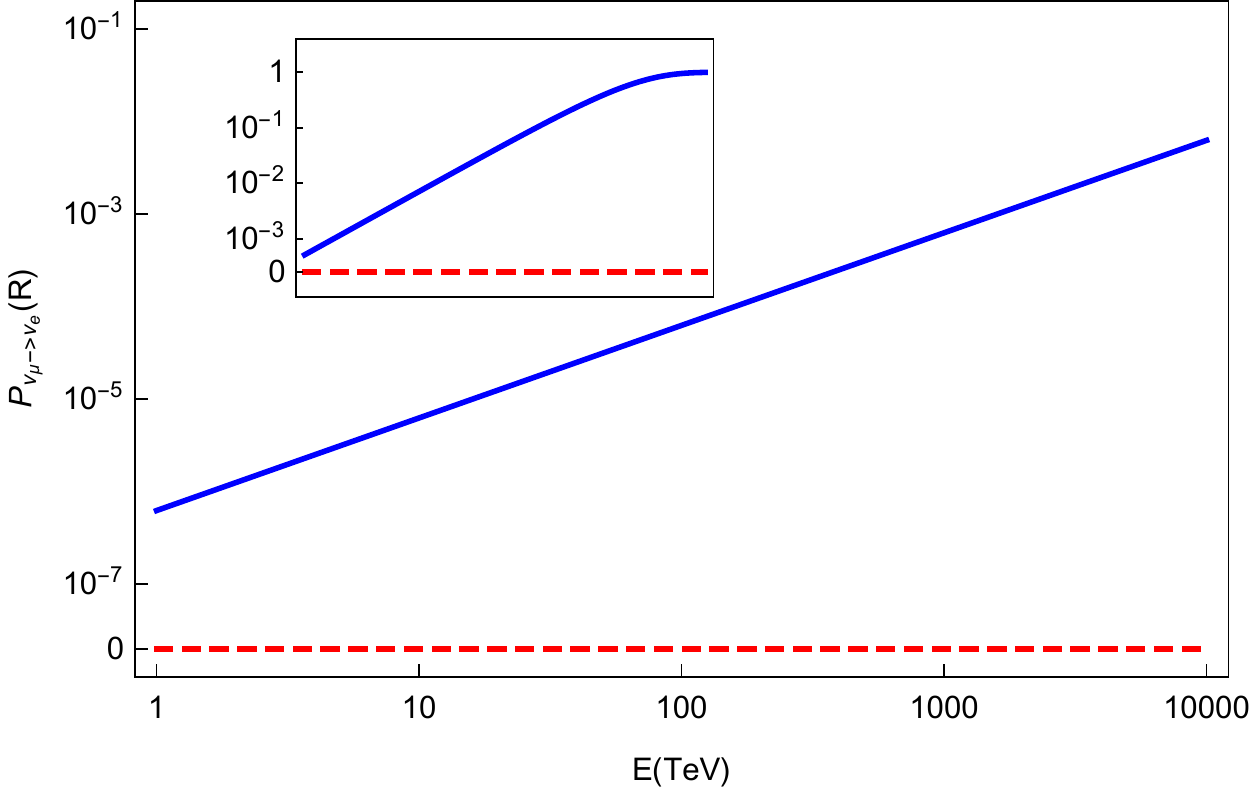}
  \includegraphics[width=0.97\linewidth]{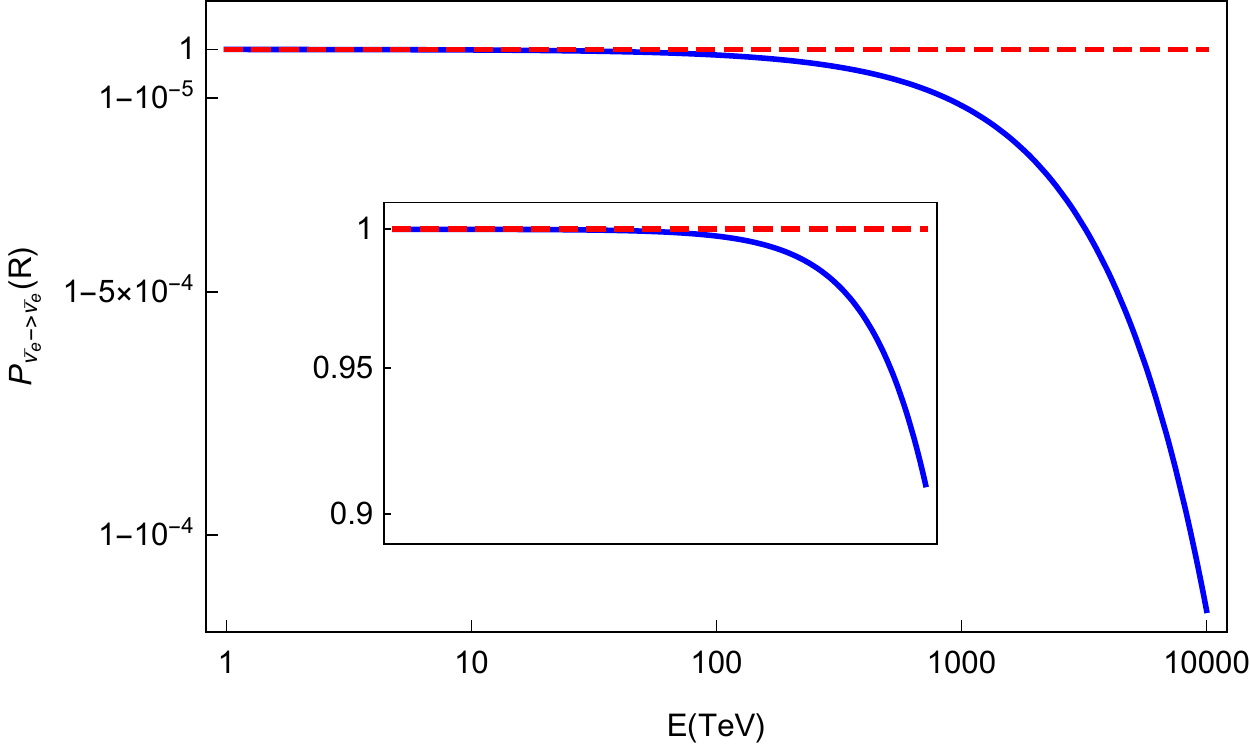}
   \caption{(color online)
Upper Panel -- Plot of the $\nu_{\mu} \rightarrow \nu_{e}$ transition probability as a function of the neutrino energy $E$, for an initial muon neutrino.
Lower Panel -- Plot of the $\bar{\nu}_{e} \rightarrow \bar{\nu}_{e}$ survival probability as a function of the antineutrino energy $E$, for an initial electron antineutrino.
In both the two plots we consider a travelling lenght equal to the diameter of the Earth. The blue solid line represents our results, while the red dashed line is obtained accordingly with the MSW Theory. For simplicity we have assumed a constant electron density $N_e = 4.069 N_A  \mathrm{cm}^{-3}$ obtained as a weighted mean from the earth density profile of the preliminary reference earth model \cite{PREM}. In the inset the same transition probabilities are plotted for a neutrino travelling through a diametral trajectory in the Sun's Core, with average electron density $N_e = 6.4 \times 10^{26} \mathrm{cm}^{-3}$ and $R_{Core} = 348000 \ \mathrm{km}$ corresponding to a quarter of the Sun's diameter.}  \label{plot_Oscillations}
\end{figure}

By contrast, when the scattering is taken into account, neutrinos still have a non--zero transition probability due to the reactions in Eq.~\eqref{AdditionalReaction}.
This point is exemplified in fig.~(\ref{plot_Oscillations}), where the transition probabilities $P_{\nu_{\beta}\rightarrow \nu_{\alpha}} (t) = Tr(\rho_{\nu_{\beta}}(t) \rho_{\nu_{\alpha}}(0))$ for neutrinos and antineutrinos in a constant high density environment is shown.
In this regime the neutrino density matrix attains a simple explicit form.
Due to the absence of muons in the medium, and since the oscillations are inhibited by the matter potential, an electron neutrino cannot ever transform in a muon neutrino.
Thus, the density matrix of an initial electron neutrino at time $t$ is simply $\rho_{\nu_e}(t) = \rho_{\nu_e}(0)$.
If instead the neutrino starts off as a muon neutrino, it can transform into an electron neutrino due to the reactions of Eq.~\eqref{AdditionalReaction}, but, for what we have just said, it can never go back to its initial state.
We can exploit this fact to give an exact prediction of the oscillation probability in this regime.
In fact, accordingly with the unitary evolution operators that we have discussed in Sec.~\ref{sec_operator}, at each single scattering with an electron, the muon neutrino has a probability $|U^{(+)}_{W,33}|^2$ of persisting in the muon state if the target electron has positive spin projection and a probability $|U^{(-)}_{W,33}|^2 = 1$ if the electron has negative spin projection. For an unpolarized medium, we expect that about a half of the electrons shall have a positive spin projection and a half a negative spin projection. At each scattering, the probability that the neutrino remains in the muonic state is then, $P_{\nu_\mu\leftrightarrow\nu_\mu} = \frac{|U^{(+)}_{W,33}|^2 + 1}{2}$. Clearly the transition probability for a single scattering is given by $P_{\nu_{\mu} \rightarrow \nu_e} = 1 - P_{\nu_\mu\leftrightarrow\nu_\mu}  = \frac{|U^{(+)}_{W,36}|^2}{2}$.
Therefore, since after a transition into an electron neutrino, it cannot transform back to the muon state, the probability that at time $t$ the neutrino is still in a muon state $(P_{\nu_\mu\leftrightarrow\nu_\mu}(t))$ is equal to
\begin{equation}
\label{Oscillation1}
P_{\nu_\mu\leftrightarrow\nu_\mu}(t)=\sum_{k=0}^{\infty} P_k(t) \left(\frac{1+|U_{W,33}|^2}{2}\right)^{k}
\end{equation}
Here $P_k(t)$ is the probability that, after a time $t$, the neutrino had $k$ scattering processes.
Now  $P_k(t)$ can be derived from the Poisson distribution with expectation value $\lambda$, where $\lambda$ is the average number of collisions occurring in a time $t$.
Denoting by $\sigma_T$ the total cross section for all the reactions of Eq.~\eqref{ParadigmReaction}, the neutrino mean free path is $l_F = \frac{1}{N_e \sigma_T}$.
The average number of collisions is simply the ratio between the distance covered by the neutrino $z$ (that coincide with $t$ assuming $c=1$) and the neutrino mean free path $\lambda = \frac{t}{l_F} = N_e \sigma_T t$.
Thus $ P_k(t) = \frac{1}{k!} (N_e \sigma_T t)^k  e^{-  N_e \sigma_T t}$ and the probability in Eq. \eqref{Oscillation1} $P_{\nu_\mu\leftrightarrow\nu_\mu}(t)$ becomes
\begin{eqnarray}
\label{Oscillation2}
P_{\nu_\mu\leftrightarrow\nu_\mu}(t)&=&\sum_{k=0}^{\infty} \frac{1}{k!} \left(\frac{N_e \sigma_T t \left(1+|U_{W,33}|^2\right)}{2}\right)^k   e^{-  N_e \sigma_T t}  \nonumber \\
&=&e^{- (1- |U_{W,33}|^{2})  \frac{N_e \sigma_T t}{2}}
\end{eqnarray}
Where in the final expression of Eq. \eqref{Oscillation2} we have recognized the Taylor expansion of the exponential. From Eq.~\eqref{Oscillation2} we immediately deduce the transition probability

\begin{equation}\label{Oscillation3}
  P_{\nu_{\mu}\rightarrow \nu_{e}} (t) =1-e^{- (1- |U_{W,33}|^{2})  \frac{N_e \sigma_T t}{2}} \ .
\end{equation}
In the same way it is possible to prove that, while a muon anti-neutrino with high energy/in high density medium does not oscillate, for an electron anti-neutrino the oscillation probability becomes
\begin{equation}\label{Oscillation4}
  P_{\bar{\nu}_{\mu}\rightarrow \bar{\nu}_{e}} (t) =1-e^{- (1- |\bar{U}_{W,11}|^{2})  \frac{N_e \sigma_T t}{2}} \ .
\end{equation}
Eq.~\eqref{Oscillation3} and Eq.~\eqref{Oscillation4} are new oscillation formulae for neutrinos and anti-neutrinos propagating through dense matter and represent the main analytical result of our paper.

We remark that in the high density/high energy limit, i. e. for $G_F N_e \gg \omega_0$ and for the kind of reactions considered, our results are compatible with those obtained via the kinetic equation approach (see for example \cite{Raffelt1993}). In order to show that, we consider Eq. (4.3) of the reference \cite{Raffelt1993}. It can be schematically written as
\begin{equation}\label{kineticEquation}
 \partial_t \rho_{\pmb{p}} = - i\sqrt{2} G_F N_e \left[G, \rho_{\pmb{p}} \right] + C[\rho_{\pmb{p}}] \ ,
\end{equation}
where $C[\rho_{\pmb{p}}]$ takes into account all the possible collisions with the particles in the medium and is given by
{\small
\begin{eqnarray*}
 && C[\rho_{\pmb{p}}] = \\
 && \frac{1}{2} \int d^3 p \Bigg[W(p',p)(1- \rho_{\pmb{p}})G \rho_{\pmb{p'}} G 
 - W(p,p') \rho_{\pmb{p}}G(1- \rho_{\pmb{p'}})G \\
 &&+ W(-p',p)(1- \rho_{\pmb{p}})G(1-\bar{\rho}_{\pmb{p'}})G - W(p,-p')\rho_{\pmb{p}}G\bar{\rho}_{\pmb{p'}}G \Bigg]  \ .\\ 
\end{eqnarray*}}
Out of all the terms, the first and the second are respectively gain and loss term due to transitions from and to other states, while the third and the fourth describe neutrino pair production and annihilation processes. We neglect the latter in our tractation right from the offset, because we are interested in generalizing the MSW effect, and thus consider only reactions that preserve the number of neutrinos. In addition,the flavor preserving reactions, as per our assumptions, affect all the flavor states only for an irrelevant common phase factor (since there is no momentum change). 
In our setting, once the initial neutrino energy is fixed, only two momenta are involved $p_{\nu_{\alpha}}, p_{\nu_{\beta}}$, so that the collision term only connects the states $\rho_{p_{\nu_{\alpha}}}$ and $\rho_{p_{\nu_{\beta}}}$.  Moreover the only possible flavor changing transition is $\nu_{\mu} + e^{-} \rightarrow \nu_{e} + \mu^{-}$, (because there are no free muons in the medium by assumption), so that the collision term actually can only connect the states $\rho_{p_{\nu_{\mu}}}$ and $\rho_{p_{\nu_e}}$, with the momenta related by momentum conservation. This implies that the time variation due to collisions is 
\begin{equation}\label{kineticEquation2} \partial_{t}\rho_{p_{\nu_{\mu}}} (t) = - W(p_{\nu_{\mu}}, p_{\nu_e}) (1- \rho_{p_{\nu_{\mu}}}(t)) \rho_{ p_{\nu_e}}
\end{equation}
and similar, but with the opposite sign, for $\rho_{\nu_e}$. Here $W(p_{\nu_{\mu}}, p_{\nu_e})$ is the transition rate for the reaction $\nu_{\mu} + e^{-} \rightarrow \nu_{e} + \mu^{-}$, which for the kinematics considered is the same rate appearing in Eq. \eqref{Oscillation2}. Finally the refractive terms (the commutator term of Eq. \eqref{kineticEquation} ) and the free oscillation term have the only effect, when combined, to inhibit the oscillations in this regime. Then the time evolution of the muon neutrino state and the related oscillation probabilities resulting from Eq. \eqref{kineticEquation2} coincide with the ones found above with our method.
Whereas this discussion proves that our method and the kinetic equation approach lead to comparable results for the approximation considered, it is worth noting that the two approaches do not necessarily agree in other circumstances.

\subsection{Decoherence}
\label{subsec_decoherence}

The charged lepton that is involved in the scattering process with the neutrino actually belongs to a thermal bath and soon after the scattering event, it starts to interact with the other particles in the medium.
Therefore, after the scattering, any information on the individual charged lepton state that emerges from the scattering is lost, whereas the neutrino state keeps evolving under the action of the mixing Hamiltonian.
These considerations can be made into a quantitative statement about the neutrino state.
For the sake of simplicity, let us take into account the case that we have considered in the previous subsection, i.e. when the high electronic density/high neutrino energy condition holds.
In such a case, as we have seen, due to the lack of oscillations, the density matrix of an electron neutrino does not change in time and  also its purity is unaffected.

A completely different scenario appears when we consider a muon neutrino.
In such a case, due to the absence of oscillation between two successive scattering processes, the purity can be written as a function of the oscillation probabilities i.e.
\begin{eqnarray}
\label{Impurity}
\mathrm{Tr}(\rho^2_\mu(t))&=&\left(P_{\nu_\mu\leftrightarrow\nu_\mu}^2(t)+P_{\nu_\mu\leftrightarrow\nu_e}^2(t) \right) \nonumber \\
&=&1-2P_{\nu_\mu\leftrightarrow\nu_\mu}(t)P_{\nu_\mu\leftrightarrow\nu_e}(t)\nonumber \\
&\le&1
\end{eqnarray}

That is, in general, lesser than one and reaches its minimum (equivalently, the impurity $1 - \mathrm{Tr}(\rho^2_\mu(t))$ reaches its maximum) when $P_{\nu_\mu\leftrightarrow\nu_\mu}(t)=P_{\nu_\mu\leftrightarrow\nu_e}(t)=\frac{1}{2}$

The inequality \eqref{Impurity} describes the loss of purity of the neutrino state due to weak-interaction scattering processes.

We conclude this subsection with an important remark.
It could be argued that the charged lepton, being part of the thermal bath, was not in a pure state, to begin with.
While this is true in general, the reasoning that leads to the inequality \eqref{Impurity} is unaffected, because it specifically pertains to the quantum correlations between the neutrino and the charged lepton.
It is the latter that gets destroyed by the interactions with the medium, leading to a loss of purity in the neutrino state, regardless of the initial charged lepton state.

\subsection{Numerical Results}
\label{subsec_comparison}

Before illustrating some numerical results let us discuss a crucial point.
The criterion for the breakdown of the matter potential approximation is represented by the high density/high energy condition $G_F N_e\gg\omega_0$).

In main sequence stars, with electron densities comparable to that of the Sun $N_{\odot}$, the matter potential approximation breaks down only at extremely high energies ($E > 100 GeV$).
On the other hand, when much denser objects are considered, such as white dwarfs with $N_{WD} \simeq 10^6 N_{\odot}$, the deviation from the matter potential approximation can be already evident at energies as low as a few $GeV$.
However, the neutrino spectrum for these objects extends from $100$ keV to $10$ MeV \cite{SolarNeutrinoSpectrum}, and so it is highly improbable that neutrinos produced within these stars reach the required energy to see appreciable modifications with respect to the MSW.

Moreover, within the core of a supernova, due to the high electron densities, a discrepancy with respect to the bare MSW prediction is to be expected for high energy neutrinos $E > 200 MeV$.
The implications of this effect on the mechanism of supernova explosions require a detailed and careful analysis, which is beyond the scope of this paper.

On the other hand, the high density/high energy condition is well verified for ultra high energy neutrinos (see for instance Ref.~\cite{Madsen2019}) propagating through the Earth that have been recently detected in the IceCube experiment\cite{Icecube}.
For this reason in our numerical analysis we will focus on them.

In  Fig.(\ref{plot_Oscillations}) we plot  $P_{\nu_{\mu} \rightarrow \nu_{e}}$ and $P_{\bar{\nu}_{e} \rightarrow \bar{\nu}_{e}}$ for ultra high energy neutrinos propagating through the Earth as a function of the neutrino energy $E$.
For sake of simplicity we consider neutrinos traveling on a diametral trajectory through the Earth and a constant electron density $N_e = 4.069 N_A  cm^{-3}$ obtained as a weighted mean from the earth density profile of the preliminary reference earth model~\cite{PREM}.
In the plots, the bare matter potential prediction (MSW effect and oscillations inhibited) (red dashed line) and the result obtained from Eq.~\eqref{Oscillation3} and Eq.~\eqref{Oscillation4} are compared.
We can see that while the MSW formula does not predict any relevant oscillation probability, our formulae provide a transition probability that is remarkably different from zero in the energy range $1-10^4$ TeV.
These predicted oscillations could be detected in future experiments. We recognize that in the energy range considered the transmission probability might be substantially lowered depending on the angle. Although this does not entirely rule out a possible observation of the phenomenon described, especially below $100  \ TeV$ where the transmission probability is fairly high, it constitutes an important limitation to the  possible experimental test of our results \cite{Aartsen2017}. Other limitations may be represented by backgrounds, detector systematic and the atmospheric
flux that are not negligible in this energy range.

\begin{figure}
\centering
  \includegraphics[width=\linewidth]{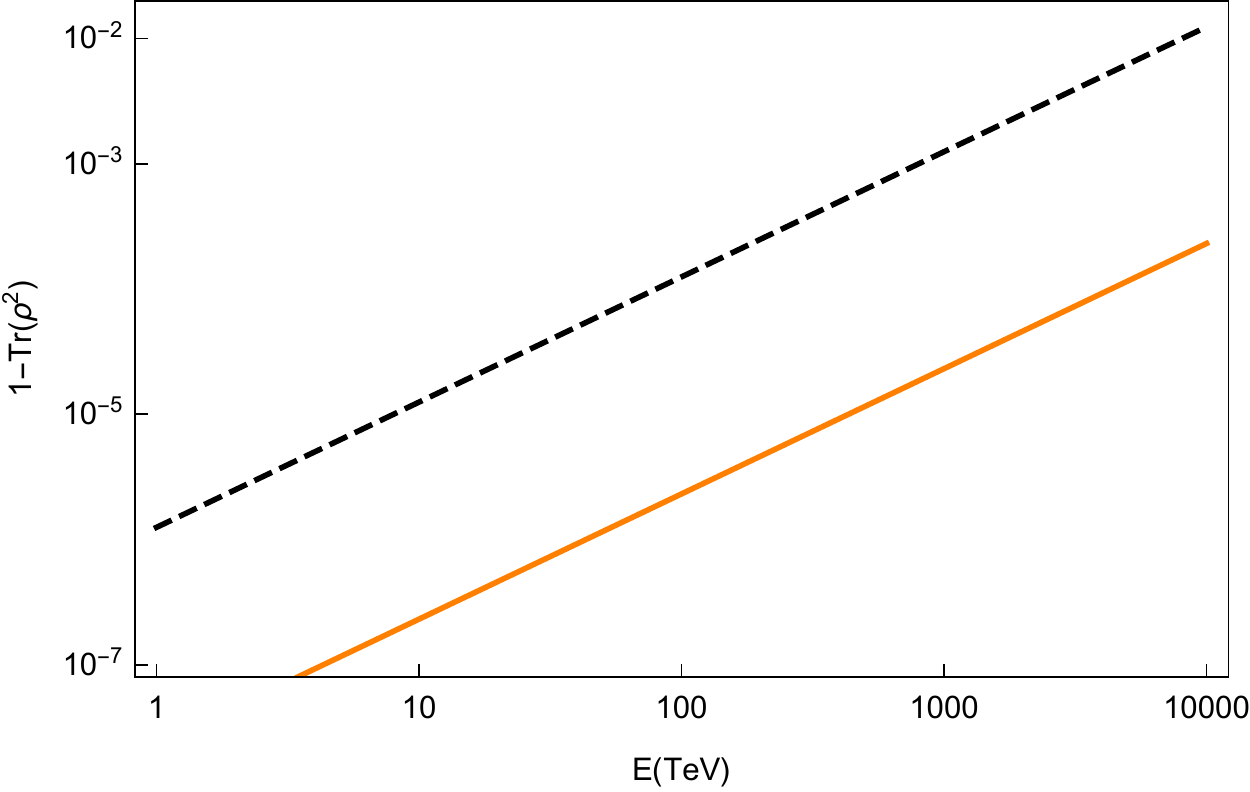}
  \caption{(color online)Plot of the impurity $1 - Tr(\rho^2)$  as a function of the neutrino energy $E$, for a neutrino travelling on a diametral trajectory through the Earth.
  The black dashed line refers to the muon neutrino state, whereas the orange solid line refers to the electron antineutrino state. We have assumed the same parameters as in fig.~(\ref{plot_Oscillations}).} \label{Impurezze}
\end{figure}

\begin{figure}[b]
\centering
  \includegraphics[width=\linewidth]{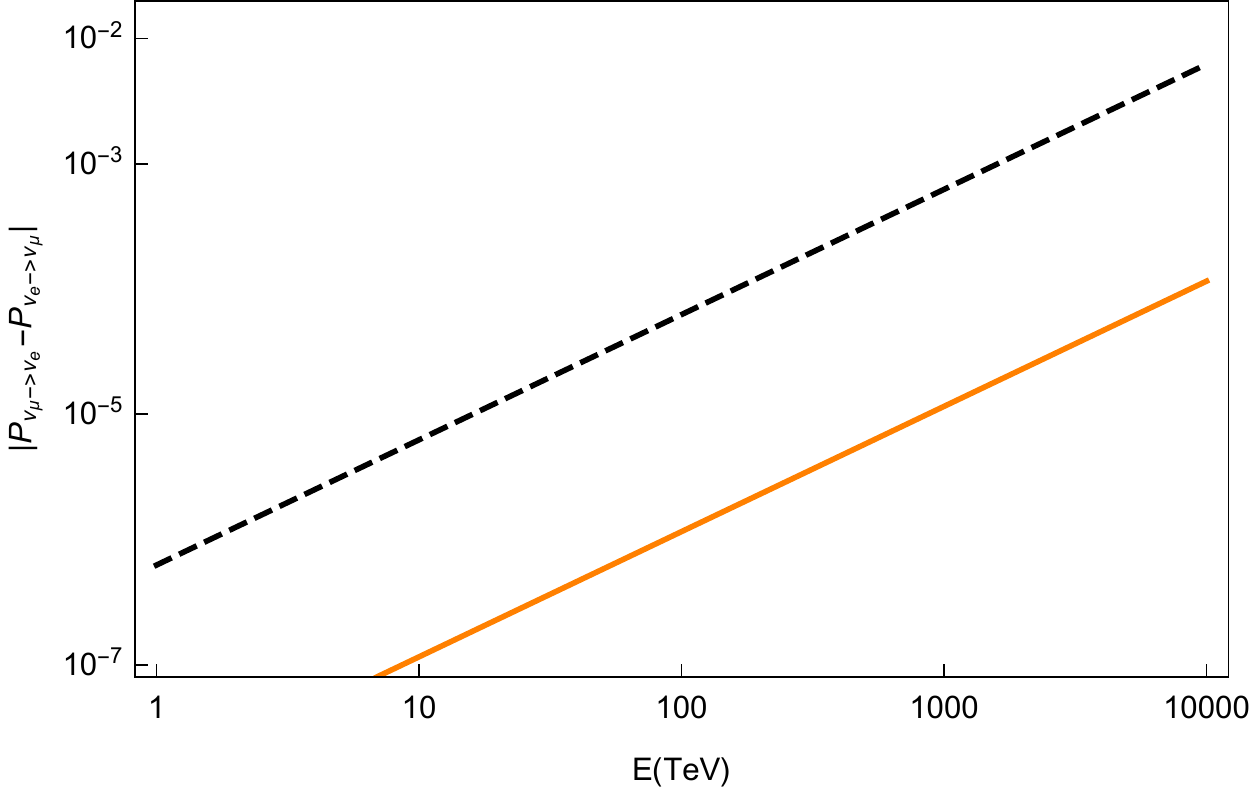}
  \caption{(color online) Plot of the $T$ asymmetries  $\Delta_T^{\mu \rightarrow e} = P_{\nu_{\mu} \rightarrow \nu_{e}} - P_{\nu_{e} \rightarrow \nu_{\mu}}$ (black dashed line) and $\bar{\Delta}_T^{\mu \rightarrow e} = P_{\bar{\nu}_{\mu} \rightarrow \bar{\nu}_{e}} - P_{\bar{\nu}_{e} \rightarrow \bar{\nu}_{\mu}}$ (orange solid line) as a function of the neutrino energy $E$, for a neutrino travelling on a diametral trajectory through the Earth.
The same parameters as in fig.~(\ref{plot_Oscillations}) are assumed.} \label{DeltaT}
\end{figure}

Also the impurity $1 - Tr(\rho^2)$ of the neutrino state $\rho$, which, as anticipated, is always zero in the matter potential approximation gain a non-trivial behavior due to the presence of the scattering processes.
In fig.~(\ref{Impurezze}) we plot the impurity  for an initial electron antineutrino and an initial muon neutrino propagating through the Earth, as a function of energy.
As can be seen from fig.~(\ref{Impurezze}), the impurity of the neutrino state initially grows with the energy.
At a specific value of the energy $E^*$, depending on the electron density and on the distance $R$ travelled in the medium, the neutrino state reaches the maximum impurity $ 1 - Tr(\rho_{\nu}^2(R)) = \frac{1}{2}$ when $P_{\nu_\mu\leftrightarrow\nu_\mu}(t)=P_{\nu_\mu\leftrightarrow\nu_e}(t)$.
Beyond $E^*$ the collisions become so frequent that the initial neutrino is almost certainly converted into a neutrino with opposite flavor, approaching the related pure state as the energy grows.

Furthermore, also the asymmetries $\Delta_T^{\alpha \rightarrow \beta} = P_{\nu_{\alpha} \rightarrow \nu_{\beta}}(t) - P_{\nu_{\beta} \rightarrow \nu_{\alpha}}(t)$ and $\Delta_{CP}^{\alpha \rightarrow \beta} = P_{\nu_{\alpha} \rightarrow \nu_{\beta}}(t) - P_{\bar{\nu}_{\alpha} \rightarrow \bar{\nu}_\beta}(t)$ are also considerably affected by the evolution in a dense environment(see Fig.\ref{DeltaT})
Indeed, in view of the possible reactions and due to the absence of muons in the medium, the $\nu_{\mu} \rightarrow \nu_{e}$ transition probability acquires a non--zero value, while the probability for the opposite transition $\nu_{e} \rightarrow \nu_{\mu}$ vanishes.
Similarly, the antineutrino transition $\bar{\nu}_{e} \rightarrow \bar{\nu}_{\mu}$ is allowed, while the opposite transition $\bar{\nu}_{\mu} \rightarrow \bar{\nu}_{e}$ is suppressed.
Overall, since $\Delta_T^{\alpha \rightarrow \beta} \neq \Delta_{CP}^{\alpha \rightarrow \beta}$, the $CPT$ symmetry is violated.
The $CPT$ asymmetry is due to the additional scattering term and the related decoherence effect.
This is in contrast with the matter potential approximation, for which, in the analyzed energy regime, $CP$, $T$ and  $CPT$ violations would all vanish identically, since the oscillations are inhibited for the MSW effect. The $CPT$ violation due to the matter effect discussed here may have played an important role during the first phases of the universe, where high densities and high neutrino energies are to be expected. 
It is worth mentioning that all the effects discussed in this section are in principle significantly enhanced when the medium electron density $N_e$ is larger, as compared to the relatively small deviations that one has for the Earth electron density. An indication of the larger deviation is shown in the insets of Figure (1), where the transitions for a high energy (anti-)neutrino travelling in the Sun's core are shown. We stress that such high energy neutrinos are not produced within the Sun, and the plot serves merely as an indication of what one expects for high energy neutrinos in denser environments. 

\section{Conclusions}
\label{sec_conclusion}

In the MSW theory, the presence of the medium is described by an effective potential that modifies the Pontecorvo Hamiltonian and the oscillation frequencies.
The term introduced in the MSW theory acts as an effective field and does not affect the property of unitarity of the evolution.
 Thus, MSW effect does not provide phenomena as decoherence.

 In the present work we  take into account the effects of the scattering processes of the neutrinos with the charged leptons in the medium.
To describe this interaction we have introduced a Hamiltonian and a unitary operator that can model the different scattering processes. Then we have connected their elements with the cross-sections derived in the context of quantum field theory.

Our approach is rather general and can provide predictions for a very wide class of physical situations.
Possible limitations of the present approach are due to the complexity of the mathematical equations. This implies that a detailed analysis is possible only in some particular cases.
Among them, two different situations have attracted our interest: 1) the case in which the probability for a single neutrino to have two or more scattering processes can be neglected,  and 2) the high-energy neutrino/high-density medium case in which $G_F N_e \gg \omega_0$.
We have shown that, in the first case the predicted results are very close to those of the MSW effect, while in the second appears a clear discrepancy between our results and the MSW ones.
Indeed, in this second regime, the new oscillation formulae that we have developed for the muon neutrino and for the electronic antineutrino show an oscillation strongly dependent on the energy, in a region in which, for the MSW theory, no oscillation is allowed.
Associated with this new oscillation formulae, we have  predicted a violation of the T, CP, and CPT theory as well as a loss of purity in the neutrino state.

A possible experimental test of our results could be made in future experiments probing neutrinos travelling through Earth at energies of the order of the $TeV$, where there is a   difference between our prediction and those of the MSW theory.

We remark that that, having neglected all the processes involving the emission or absorption of neutrinos, the dissipative evolution we obtain for neutrinos in a medium cannot be precisely compared with the gain-term and the loss-term structure obtained in various papers, as for instance in \cite{BaryonAsymmetry} in a different context. A further analysis of the neutrino propagation in a dense medium, including creation and destruction processes, shall be treated in a forthcoming paper.

For the sake of simplicity, we have developed our results in the framework of quantum mechanics.
Obviously, the inclusion of field--theoretical effects can be of interest even if
the QFT effects on particle mixing and oscillations \cite{Blasone:1998hf, Capolupo:2006et} and the curvature effects on neutrino oscillations \cite{Capolupo2020Curv} are negligible.

\section*{Acknowledgements}

A.C. and A.Q. thank the partial financial support from MIUR and INFN. A.C. also thanks the COST Action CA1511 Cosmology and Astrophysics Network for Theoretical Advances and Training Actions (CANTATA).
S.M.G. acknowledges support from the European Regional Development Fund the Competitiveness and Cohesion Operational Programme (KK.01.1.1.06--RBI TWIN SIN) and the Croatian Science Fund Project No. IP-2016--6--3347 and IP-2019--4--3321.
S.M.G. also acknowledges the QuantiXLie Center of Excellence,  co--financed by the Croatian Government and European Union  (Grant KK.01.1.1.01.0004).

\appendix

\section{Scattering amplitudes}\label{FirstAppendix}

In this appendix we derive the scattering amplitudes for the various processes involved. We shall study the reactions in the rest frame of the initial charged lepton and assume that the $4$-momentum exchange is small compared to the mass of the weak gauge bosons $m_{W},m_{Z}$. The weak boson propagator in momentum space is given by
\begin{equation}\label{WeakPropagator}
 D_{\mu \nu}^j (q) = \frac{-i \left(g_{\mu \nu} - \frac{q_{\mu} q_{\nu}}{m^2_j}\right)}{q^2 - m_j^2 + i\epsilon}
\end{equation}
with $j=W,Z$. When $q^2 \ll m_j^2$, we can neglect all the $q$ terms in Eq.\eqref{WeakPropagator}. Correspondingly the weak propagator becomes a constant, and the weak interaction is replaced by the effective Fermi current-current interaction. There are three kinds of reactions to consider: neutral current interactions like $\nu_{e} + \mu \leftrightarrow \nu_{e} + \mu$, charged current interactions like $\nu_e + \mu \leftrightarrow \nu_{\mu} + e$ and reactions where both charged and neutral current contribute, like $\nu_{e} + e \leftrightarrow \nu_{e} + e$.
We start by computing $\mathcal{M}(\nu_{e} + e \leftrightarrow \nu_{e} + e)$. We work in the rest frame of the initial electron, where $p_{e} \equiv (m_e,\pmb{0})$ and $p_{\nu} \equiv (E_{\nu},\pmb{p_{\nu}})$. The hypothesis of vanishing $4$-momentum exchange implies that $p'_{\nu}= p_{\nu}$ and $p'_e=p_e$. Without loss of generality, we align the $z$ axis with the neutrino $3$-momentum $\pmb{p_{\nu}} = \pmb{p'_{\nu}}$. It is convenient to work in the helicity basis. Here the Dirac matrices are
\begin{equation}
 \gamma^0 = \begin{pmatrix} 0 & -\mathbb{1} \\ - \mathbb{1}& 0\end{pmatrix} \ \ \ \pmb{\gamma} = \begin{pmatrix} 0 & \pmb{\sigma} \\ - \pmb{\sigma} & 0\end{pmatrix} \ \ \ \gamma_5  = \begin{pmatrix} \mathbb{1} & 0 \\ 0 & -\mathbb{1} \end{pmatrix}
\end{equation}

with $\mathbb{1}$ the $2\times2$ identity matrix and $\pmb{\sigma}$ the vector of Pauli matrices. In this basis the spinors with helicity $h = \pm$ are
\begin{eqnarray}
\nonumber u^{(h)}(\pmb{p}) &=& \begin{pmatrix} - \sqrt{E_{\pmb{p}} + h |\pmb{p}|} \chi^{(h)} (\pmb{p}) \\ \sqrt{E_{\pmb{p}} - h |\pmb{p}|} \chi^{(h)} (\pmb{p}) \end{pmatrix} \\
\nonumber v^{(h)}(\pmb{p}) &=& -h\begin{pmatrix} - \sqrt{E_{\pmb{p}} - h |\pmb{p}|} \chi^{(-h)} (\pmb{p}) \\ \sqrt{E_{\pmb{p}} + h |\pmb{p}|} \chi^{(-h)} (\pmb{p}) \end{pmatrix}
\end{eqnarray}
respectively for the particle and the antiparticle. The two-spinors $\chi$ depend upon the orientation of $\pmb{p}$. When $\pmb{p}$ is along the $z$ axis or $\pmb{p} = 0$ (this is the only case relevant to us), we have $\chi^{(+)}= \begin{pmatrix} 1 \\ 0 \end{pmatrix}$ and $\chi^{(-)} =  \begin{pmatrix} 0 \\ 1 \end{pmatrix}$. The charged current contribution is given by \cite{GiuntiKim}
\begin{eqnarray}
 \nonumber \mathcal{M}_{CC} &=& - \frac{G_F}{\sqrt{2}} J^{\mu \dagger}_{\nu_e e} J_{\mu \nu_e e} \\ &=& - \frac{G_F}{\sqrt{2}} \left(\bar{e} \gamma^{\mu} (1-\gamma_5)\nu_e \right) \left(\bar{\nu}_e \gamma_{\mu} (1-\gamma_5)e \right) \ .
\end{eqnarray}
Only left-handed neutrinos participate in the weak interaction, whereas both the spin orientations must be considered for the electron. We then have
\begin{eqnarray}
 \nonumber J^{\mu \dagger}_{\nu_e e (h) } &=& 2\overline{u^{(h)}}(p'_e) \gamma^{\mu} P_L u^{(-)}(p_{\nu}) \\
 J^{\mu}_{\nu_e e (h)} &=& 2 \overline{u^{(-)}}(p'_{\nu}) \gamma^{\mu} P_L u^{(h)}(p_e)
\end{eqnarray}
where we have introduced the left-handed projector $P_L= (1-\gamma_5)/2$ for convenience. The currents can be easily computed, giving
\begin{eqnarray*}
 J^{\mu \dagger}_{\nu_e e (+)} &=& 2\sqrt{(E'_e - |\pmb{p'_e}|)(E_{\nu}+|\pmb{p_{\nu}}|}) \chi^{(+) \dagger}(\pmb{p'_e})\bar{\sigma}^{\mu} \chi^{(-)}(\pmb{p_{\nu}}) \\
 J^{\mu \dagger}_{\nu_e e (-)} &=& 2\sqrt{(E'_e + |\pmb{p'_e}|)(E_{\nu}+|\pmb{p_{\nu}}|}) \chi^{(-) \dagger}(\pmb{p'_e})\bar{\sigma}^{\mu} \chi^{(-)}(\pmb{p_{\nu}}) \\
 J^{\mu }_{\nu_e e (+)} &=& 2\sqrt{(E_e - |\pmb{p_e}|)(E'_{\nu}+|\pmb{p'_{\nu}}|}) \chi^{(-) \dagger}(\pmb{p'_{\nu}})\bar{\sigma}^{\mu} \chi^{(+)}(\pmb{p_{e}}) \\
 J^{\mu }_{\nu_e e (-)} &=& 2\sqrt{(E_e + |\pmb{p_e}|)(E'_{\nu}+|\pmb{p'_{\nu}}|}) \chi^{(-) \dagger}(\pmb{p'_{\nu}})\bar{\sigma}^{\mu} \chi^{(-)}(\pmb{p_{e}})
\end{eqnarray*}
where $\bar{\sigma}^{\mu} \equiv (\mathbb{1},-\pmb{\sigma})$. Considered that $\pmb{p_{\nu}}=\pmb{p'_{\nu}} \ || \ z$, $|\pmb{p}_{\nu}| \simeq E_{\nu} $ and that $\pmb{p_e} = \pmb{p'_e} = \pmb{0}$, we have
\begin{eqnarray*}
 J^{\mu \dagger}_{\nu_e e (+)} &\equiv& -2 \sqrt{2E_{\nu} m_e} (0,1,-i,0) \\
  J^{\mu \dagger}_{\nu_e e (-)} &\equiv& 2 \sqrt{2E_{\nu} m_e} (1,0,0,1) \\
  J^{\mu }_{\nu_e e (+)} &\equiv& -2 \sqrt{2E_{\nu} m_e} (0,1,i,0) \\
  J^{\mu }_{\nu_e e (-)} &\equiv& 2 \sqrt{2E_{\nu} m_e} (1,0,0,1)  \ .
\end{eqnarray*}
It follows immediately that the charged current contribution is zero for opposite initial and final electron spin projections. On the other hand we find
\begin{equation}
 \mathcal{M}_{CC}^{++} = 8\sqrt{2} G_F E_{\nu} m_e  \ \ \ \ \mathcal{M}_{CC}^{--} = 0   \ ,
\end{equation}
Where $\mathcal{M}^{hh'}$ denotes the amplitude associated to the charged lepton with initial helicity $h$ and final helicity $h'$.
The neutral current contribution for the same process is \cite{GiuntiKim}
\begin{equation}
 \mathcal{M}_{NC} = - \frac{G_F}{\sqrt{2}} \left( \overline{\nu_e} \gamma^{\mu}(1-\gamma_5) \nu_e \right) \left(\bar{e}\gamma_{\mu}(g^e_V-g^e_A\gamma_5) \right)
\end{equation}
with $g^e_V = -\frac{1}{2} + \sin^2 \theta_W$ and $g^e_A=- \frac{1}{2}$. The calculation follows the same steps as before, yielding $\mathcal{M}_{NC}^{+-} = 0 = \mathcal{M}_{NC}^{-+}$ and
\begin{eqnarray}
 \nonumber \mathcal{M}_{NC}^{++} &=& -8\sqrt{2}G_F E_{\nu}m_e \left(\sin^2 \theta_W - \frac{1}{2}\right) \\
 \mathcal{M}_{NC}^{--} &=& -8\sqrt{2}G_F E_{\nu}m_e \sin^2 \theta_W \ .
\end{eqnarray}
The total amplitude for the process is therefore
\begin{widetext}
\begin{eqnarray}\label{FirstAmplitude}
 \nonumber \mathcal{M}^{++} (\nu_e + e^{-} \leftrightarrow \nu_e + e^{-}) &=& -8\sqrt{2}G_F E_{\nu}m_e\left(\sin^2 \theta_W + \frac{1}{2}\right) \\
 \mathcal{M}^{--}(\nu_e + e^{-} \leftrightarrow \nu_e + e^{-}) &=& -8\sqrt{2}G_F E_{\nu}m_e \sin^2 \theta_W \ .
\end{eqnarray}
The situation is similar for the other process in which both neutral and charged current contribute, i.e. $\nu_{\mu} + \mu \leftrightarrow \nu_{\mu} + \mu$:
\begin{eqnarray}
 \nonumber \mathcal{M}^{++} (\nu_{\mu} + \mu^{-} \leftrightarrow \nu_{\mu} + \mu^{-})&=& -8\sqrt{2}G_F E_{\nu}m_{\mu}\left(\sin^2 \theta_W + \frac{1}{2}\right) \\
 \mathcal{M}^{--} (\nu_{\mu} + \mu^{-} \leftrightarrow \nu_{\mu} + \mu^{-})&=& -8\sqrt{2}G_F E_{\nu}m_{\mu} \sin^2 \theta_W \ .
\end{eqnarray}
For the elastic scatterings $\nu_e + \mu \leftrightarrow \nu_e + \mu$ and $\nu_{\mu} + e \leftrightarrow \nu_{\mu} + e$ only the neutral current contributes. We have
\begin{eqnarray}
 \nonumber \mathcal{M}^{++} (\nu_e + \mu^{-} \leftrightarrow \nu_e + \mu^{-}) &=& -8\sqrt{2}G_F E_{\nu}m_{\mu}\left(\sin^2 \theta_W - \frac{1}{2}\right) \\
 \mathcal{M}^{--} (\nu_e + \mu^{-} \leftrightarrow \nu_e + \mu^{-}) &=& -8\sqrt{2}G_F E_{\nu}m_{\mu} \sin^2 \theta_W
\end{eqnarray}
and
\begin{eqnarray}
 \nonumber \mathcal{M}^{++} (\nu_{\mu} + e^{-} \leftrightarrow \nu_{\mu} + e^{-}) &=& -8\sqrt{2}G_F E_{\nu}m_{e}\left(\sin^2 \theta_W - \frac{1}{2}\right) \\
 \mathcal{M}^{--} (\nu_{\mu} + e^{-} \leftrightarrow \nu_{\mu} + e^{-})&=& -8\sqrt{2}G_F E_{\nu}m_{e} \sin^2 \theta_W
\end{eqnarray}
respectively.
The quasi-elastic scattering $\nu_{\mu} + e \leftrightarrow  \nu_{e} + \mu$ is the only process considered that has an energy threshold and cannot take place with vanishing $4$-momentum exchange. Calling $p_{\nu}, p_{e}$ the initial $4$-momenta and $p'_{\nu},p_{\mu}$ the final $4$-momenta, momentum conservation implies
\begin{eqnarray*}
 m_e + E_{\nu} &=& E'_{\nu} + E_{\mu} \\
 \pmb{p_{\nu}} &=& \pmb{p'_{\nu}} + \pmb{p_{\mu}}
\end{eqnarray*}
as expressed in the rest frame of the initial electron. In order that the neutrino is forward scattered $\pmb{p_{\nu}} \ || \ \pmb{p'_\nu}  \ || \ z$, all the three-momenta must lie on the $z$ axis. Projecting the second equation on this axis and recalling that $E_{\nu} \simeq p_{\nu}$, $E'_{\nu} \simeq p'_{\nu}$, we obtain by subtraction
\begin{equation}\label{EMConservation}
 E_{\mu} = p_{\mu} + m_e \ .
\end{equation}
 Squaring equation \eqref{EMConservation} we find the muon $4$-momentum
\begin{equation}\label{Kinematics1}
 p_{\mu} =\frac{m_{\mu}^2 - m_e^2}{2m_e} \Longrightarrow E_{\mu} = \frac{m_e^2 + m_{\mu}^2}{2m_e}
\end{equation}
and the energy of the outcoming neutrino $E'_{\nu} = E_{\nu}- \frac{m_{\mu}^2 - m_e^2}{2m_e}$. Evidently the energy threshold is set by $\frac{m_{\mu}^2 - m_e^2}{2m_e} \simeq 10  \ \mathrm{GeV}$ and the (lightlike) $4$-momentum exchange $q=p_{\nu}-p'_{\nu}$ has components of the same order $q^0 = q^z \simeq 10 \  \mathrm{GeV}$. Since $q$ is lightlike, the denominator in the propagator expression \eqref{WeakPropagator} is again a constant, but we make an error of order $\left(\frac{q^0}{m_W}\right)^2 \simeq \frac{1}{64}$ in neglecting the $q$-term in the numerator. Nonetheless we stick to the current-current interaction, and find the amplitudes as
\begin{equation}
 \mathcal{M}^{++} (\nu_{\mu} + e^{-} \rightarrow \nu_e + \mu^{-}) = 8\sqrt{2} G_F m_e \sqrt{E_{\nu}E'_{\nu}} \ \ \ \ \ \ \ \ \ \ \  \mathcal{M}^{--}(\nu_{\mu} + e^{-} \rightarrow \nu_e + \mu^{-}) = 0   \ .
\end{equation}
The calculation of the amplitudes for the antineutrinos follows the same steps, except that now only the right-handed antineutrino participates in the interaction and the quasi-elastic reactions are $\bar{\nu}_e + e \leftrightarrow \nu_{\mu} + \mu$. We have
\begin{eqnarray}
 \nonumber \mathcal{M}^{++} (\bar{\nu}_e + e^{-} \leftrightarrow \bar{\nu}_e + e^{-}) &=& 8\sqrt{2}G_F E_{\nu}m_e\left(\frac{3}{2}-\sin^2 \theta_W\right) \\
 \mathcal{M}^{--}(\bar{\nu}_e + e^{-} \leftrightarrow \bar{\nu}_e + e^{-}) &=& -8\sqrt{2}G_F E_{\nu}m_e \sin^2 \theta_W \ .
\end{eqnarray}
for the $\bar{\nu}_e + e \leftrightarrow \bar{\nu}_e + e$ scattering and
\begin{eqnarray}
 \nonumber \mathcal{M}^{++}(\bar{\nu}_{\mu} + \mu^{-} \leftrightarrow \bar{\nu}_{\mu} + \mu^{-}) &=& 8\sqrt{2}G_F E_{\nu}m_{\mu}\left(\frac{3}{2}-\sin^2 \theta_W\right) \\
 \mathcal{M}^{--}(\bar{\nu}_{\mu} + \mu^{-} \leftrightarrow \bar{\nu}_{\mu} + \mu^{-}) &=& -8\sqrt{2}G_F E_{\nu}m_{\mu} \sin^2 \theta_W \ .
\end{eqnarray}
for the $\bar{\nu}_{\mu} + \mu \leftrightarrow \bar{\nu}_{\mu} + \mu$
scattering. For the neutral current reactions one finds
\begin{equation}
 \mathcal{M}^{++} (\bar{\nu}_{\mu} + e^{-} \leftrightarrow \bar{\nu}_{\mu} + e^{-}) = 8 \sqrt{2}G_F E_{\nu} m_e \ \ \ \ \ \ \ \ \ \mathcal{M}^{--} (\bar{\nu}_{\mu} + e^{-} \leftrightarrow \bar{\nu}_{\mu} + e^{-})= 0
\end{equation}
for $\bar{\nu}_{\mu} + e \leftrightarrow \bar{\nu}_{\mu} + e$
and
\begin{equation}
 \mathcal{M}^{++}(\bar{\nu}_{e} + \mu^{-} \leftrightarrow \bar{\nu}_{e}  + \mu^{-}) = 8 \sqrt{2}G_F E_{\nu} m_\mu \ \  \ \ \ \ \ \ \ \mathcal{M}^{--}(\bar{\nu}_{e} + \mu^{-} \leftrightarrow \bar{\nu}_{e}  + \mu^{-}) = 0
\end{equation}
for $\bar{\nu}_{e} + \mu \leftrightarrow \bar{\nu}_{e}  + \mu$. Finally, the quasi-elastic scattering is described by the same kinematics of equation \eqref{Kinematics1} and its amplitude is
\begin{equation}
 \mathcal{M}^{++} (\bar{\nu}_{e} + e^{-} \rightarrow \bar{\nu}_{\mu} + \mu^{-} ) = 8\sqrt{2} G_F m_e \sqrt{E_{\nu}E'_{\nu}} \ \ \ \ \ \ \ \ \ \mathcal{M}^{--}(\bar{\nu}_{e} + e^{-} \rightarrow \bar{\nu}_{\mu} + \mu^{-} ) = 0   \ .
\end{equation}
\end{widetext}


\begin{thebibliography}{800}
\bibitem{Vannucci}
F. Vannucci, \textit{Progress in Particle and Nuclear Physics}, vol. \textbf{85}, July 2017, pp. 1-47 (2017).

\bibitem{Nakamura1}
F. P. An et al. [Daya-Bay Collaboration],
Phys. Rev. Lett. {\bf 108}, 171803 (2012)

\bibitem{Nakamura2}
 J. K. Ahn et al. [RENO Collaboration],
 Phys. Rev. Lett. {\bf 108}, 191802 (2012)

\bibitem{Nakamura3}3
 Y. Abe et al. [Double Chooz Collaboration],
Phys. Rev. Lett. {\bf 108}, 131801 (2012)

\bibitem{Nakamura4}
K. Abe et al. [T2K Collaboration],
Phys. Rev. Lett. {\bf 107}, 041801 (2011).

\bibitem{Nakamura5}
P. Adamson et al. [MINOS Collaboration],
Phys. Rev. Lett. {\bf 107}, 181802 (2011).


\bibitem{Nakamura6}
 K. Nakamura and S.T. Petcov,
Phys. Rev. D {\bf 86 }, 010001 (2012).

\bibitem{Capolupo2018}
A. Capolupo, S.M. Giampaolo, B.C. Hiesmayr, G. Vitiello,
Phys. Lett. B {\bf 780} 216, (2018).





\bibitem{Pontecorvo}
S.~M.~Bilenky and
B.~Pontecorvo,  Phys. Rep.  {\bf 41} 225; (1978)\\
S.~M.~Bilenky and S.~T.~Petcov,
{\it Rev.\ Mod.\ Phys.} {\bf 59} (1987) 671.

\bibitem{CapolupoCPTdiss}
  K. Simonov, A. Capolupo, S.M. Giampaolo,
  European Physical Journal C, {79}, 902 (2019).

\bibitem{MSW1}
S. P. Mikheev, A. Yu. Smirnov,
  Sov. J. Nuc. Phys.  {\bf 42} (6): 913, (1985).

\bibitem{MSW2}
L. Wolfenstein
  Phys. Rev.  D {\bf 17} (9): 2369, (1978).

\bibitem{Lind}
G. Lindblad,
Communications in Mathematical Physics, vol. {\bf 48}, no.
2, pp. 119, (1976);
V. Gorini, A. Kossakowski, and E. C. Sudarshan,
Journal of Mathematical Physics, vol. {\bf 17}, no. 5, p. 821, (1976).

 \bibitem{Benatti}
 F. Benatti, R. Floreanini, JHEP {\bf 02}, 32, (2000);
 Phys. Rev. D {\bf 64}, 085015 (2001);
E. Lisi, A. Marrone and D. Montanino, Phys. Rev. Lett.
{\bf 85}, 1166 (2000);
A.M. Gago, E.M. Santos, W.J.C. Teves and R.
Zukanovich Funchal, Phys. Rev. D {\bf 63}, 073001 (2001);
 D. Morgan, E. Winstanley, J. Brunner and L. F. Thompson,
Astropart. Phys. {\bf 25}, 311 (2006);
 G. L. Fogli, E. Lisi, A. Marrone, D. Montanino and A.
Palazzo, Phys. Rev. D {\bf 76}, 033006 (2007);
Y. Farzan, T. Schwetz and A.Y. Smirnov, JHEP {\bf 0807},
067 (2008).

\bibitem{Benatti1}
R.L.N. Oliveira and M.M. Guzzo, Eur. Phys. J. C {\bf 69},
493 (2010); Eur. Phys. J. C {\bf 73},
2434 (2013);
 R.L.N Oliveira, Eur. Phys. J. C {\bf 76}, 417 (2016);
 P Bakhti, Y. Farzan and T. Schwetz, JHEP {\bf 05}, 007
(2015); G. Balieiro Gomes, M. M. Guzzo, P. C. de Holanda and
R. L. N. Oliveira, Phys. Rev. D {\bf 95}, 113005 (2016);
 M.M. Guzzo, P.C. de Holanda and R.L.N. Oliveira, Nucl.
Phys. B {\bf 908}, 408 (2016).

\bibitem{CapolupoDec}
  A. Capolupo, S.M. Giampaolo, G. Lambiase,
  Phys. Lett. B, {\bf 792}, 298 (2019);
    L.~Buoninfante, A.~Capolupo, S.~M.~Giampaolo and G.~Lambiase,
  arXiv:2001.07580 [hep-ph].


\bibitem{Stodolsky1987}
L. Stodolsky, \textit{Phys. Rev. D}, \textbf{36}, 2273 (1987)

\bibitem{Raffelt1993}
G. Sigl, G. Raffelt, \textit{Nucl. Phys. B}, vol. \textbf{406}, issues 1-2, pp. 423-451 (1993).

\bibitem{Vlasenko2013}
A.~Vlasenko, G.~M. Fuller, and V.~Cirigliano, {\em Phys. Rev.} {\bf D89} (2014), no.~10 105004 (2014).

\bibitem{Blaschke2016}
D.~N. Blaschke and V.~Cirigliano, {\em Phys. Rev.} {\bf D94} (2016), no.~3 033009 (2016).

\bibitem{Raffelt2018}
T. Stirner, G. Sigl, G. Raffelt, \textit{JCAP05 (2018)016}, vol. \textbf{2018} (2018).

\bibitem{Lewis}
R. R. Lewis, \textit{Phys. Rev. D} \textbf{21}, 3 (1980).

\bibitem{GiuntiKim}
C. Giunti, C. W. Kim, \textit{Fundamentals of Neutrino Physics and Astrophysics}, pp. 322-343, Oxford University Press (2007).

\bibitem{Note1.5}
Actually the proportionality $\sigma_{\nu l} \propto E_{\nu}$ does not hold in all the reference frames. However, the total cross section is proportional to the Lorentz invariant quantity $s = (p_{\nu} + p_{l})^2$, so that it is always a monotonically increasing function of the neutrino energy.

\bibitem{Note1.625}
We remark that this is not the most general form for the state of the medium $\rho_{Medium}$, which is a generic many--particle state. Nevertheless the knowledge of the exact state of the medium is irrelevant to the purpose of determining the neutrino propagation and the associated decoherence. The only quantity related to $\rho_{Medium}$ which is relevant at this level is represented by the electron density $N_e (x)$. Complications related with the exact form of $\rho_{Medium}$ (for instance as a Fermi-Dirac distribution) are deferred to later works.

\bibitem{Note1.75}
 A homogeneous medium has a uniform electron density $N_e$, indipendent of the position $\pmb{x}$ within the medium. The collision rate $\Gamma = N_e \sigma_T$ depends only on the electron density $N_e$ and on the neutrino energy $E$, and since the latter do not change during the propagation, the former is a constant. Also, the collisions occur independently from each other, so that the probability of having a collision at a given time $t$ does not depend on the previous scattering events. These properties are compatible with a Poisson distribution.

\bibitem{Note2}
Only left--handed neutrinos and right--handed antineutrinos participate in the weak interaction processes \cite{Vannucci}.

\bibitem{PDG}
P.A. Zyla et al. (Particle Data Group), to be published in Prog. Theor. Exp. Phys. 2020, 083C01 (2020), \textit{Kinematics} (2020)

\bibitem{Shapiro}
J. Helayel-Neto, A. B. Penna-Firme, I. Shapiro, \textit{Journal of High Energy Physics}, vol \textbf{2000}, JHEP01 (2000).

\bibitem{Weinberg}
S. Weinberg, \textit{The Quantum Theory of Fields, Vol. I: Foundations}, First edition, Cambridge University Press (1995).


\bibitem{Aitchison}
I. J. R. Aitchison, A. J. G. Hey, \textit{Gauge Theories in Particle Physics}, 3rd edition, vol. 2, IOP Publishing (2004).

\bibitem{FormaggioZeller}
J. A. Formaggio, G. P. Zeller, \textit{Rev. Mod. Phys.} \textbf{84}, 1307 (2012).




\bibitem{Note4}
The charged lepton fields can be independently rephased, but neutrino fields can only be rephased in a flavor--independent manner if one whishes to keep the real form of the mixing Hamiltonian of Eq. \eqref{MixingHamiltonian}. This leaves a total of 4 phases that can be independently specified, corresponding to $\ket{e}, \ket{\mu}, \ket{\nu_{\alpha}}, \ket{\bar{\nu}_{\alpha}}$.




\bibitem{SolarNeutrinoSpectrum}
D. D'Angelo et al. EPJ Web Conf., 126 (2016) 02008 (2016).

\bibitem{Madsen2019}
 J. Madsen (for the IceCube collaboration), \textit{Ultra-High Energy Neutrinos}, 	arXiv:1901.02528 [astro-ph.HE] (2019).


\bibitem{Icecube}
 A. Ishihara and for the IceCube Collaboration 2016 \textit{J. Phys.: Conf. Ser.}, \textbf{718}, 062027 (2016).

\bibitem{PREM}
 A. Ereditato et al., \textit{The State of the Art of Neutrino Physics}, Advanced Series on Directions in High Energy Physics -- Vol. \textbf{28}, {World Scientific, 2018}.

\bibitem{Aartsen2017}
M. Aartsen, M. Ackermann et al (The IceCube Collaboration). Measurement of the multi-TeV neutrino interaction cross-section with IceCube using Earth absorption. Nature \textbf{551}, pp. 596-600 (2017).

\bibitem{BaryonAsymmetry}
J.-S. Gagnon, M. Shaposhnikov, Phys. Rev. D \textbf{83}, 065021 (2011).








%
%
%
%

\bibitem{Blasone:1998hf}
E.~Alfinito, M.~Blasone, A.~Iorio and G.~Vitiello,
  Phys.\ Lett.\ B {\bf 362}, 91 (1995);
 Ch. Y. Cardall, 
 Phys. Rev. D 61, 073006 (2000);
M.~Blasone, A.~Capolupo, G.~Vitiello,
 {\it  Phys.\ Rev.\ D} {\bf 66}, 025033 (2002) and references therein.
K. C. Hannabuss and D. C. Latimer, J. Phys. A 33, 1369 (2000);  J. Phys. A 36, L69 (2003);
%
M. Beuthe,
Phys. Rep. 375, 105 (2003);
 M. Blasone, A. Capolupo, O. Romei and  G. Vitiello,
  Phys.\ Rev.\ D  {\bf 63}, 125015  (2001).
C. R. Ji and Y. Mishchenko, Phys. Rev. D {\bf 64}, 076004 (2001);
M. Beuthe,
Phys. Rev. D {\bf 66}, 013003 (2002).
 A. Capolupo, C. R. Ji ,  Y. Mishchenko and  G. Vitiello,
  Phys.\ Lett.\ B  {\bf 594}, 135 (2004).
K. Fujii, C. Habe, and T. Yabuki, Phys. Rev. D 59, 113003 (1999); Phys. Rev. D 64, 013011 (2001);
  M.~Blasone, A.~Capolupo, F.~Terranova, G.~Vitiello,
   Phys.\ Rev.\ D {\bf 72}, 013003 (2005).
C. C. Nishi,
Phys. Rev. D {\bf 73}, 053013 (2006).
E. Kh. Akhmedov and A. Wilhelm,
J. High Energy Phys. 2013 (01), 165 (2013);
A. Kobach, A. V. Manohar, and J. McGreevy,
Phys. Lett. B {\bf 783}, 59 (2018);
Naumov, D.V., Naumov, V.A.
Phys. Part. Nuclei {\bf 51}, 1  (2020).



\bibitem{Capolupo:2006et}
  A.~Capolupo,
  Adv. High Energy Phys.  {\bf 2016},   8089142, 10 (2016);
  A. Capolupo, S. Capozziello and G. Vitiello,
   Phys.\ Lett.\ A  {\bf 373}, 601 (2009);
%
%
    Phys.\ Lett.\ A  {\bf 363}, 53 (2007);
%
  Int.\ J.\ Mod.\ Phys.\ A  {\bf 23}, 4979 (2008);
  %
   M. Blasone,  A. Capolupo,  S. Capozziello,  S. Carloni and  G. Vitiello,
  Phys.\ Lett.\ A   {\bf 323}, 182 (2004).
%
  A.~Capolupo, M.~Di Mauro, A.~Iorio,
    Phys.\ Lett.\ A  {\bf 375}, 3415 (2011).
%
%
A.~Capolupo, G.~Lambiase, A.~Quaranta
 Phys.\ Rev.\ D {\bf 101}, 095022 (2020).
%
%
A. Capolupo, G. Lambiase, A. Quaranta, S.M. Giampaolo,
Eur. Phys. J. C, {\bf 80}, 423 (2020).
%
%
  A. Capolupo, I. De Martino, G. Lambiase and An. Stabile,
  Phys. Lett. B, {\bf 790}, 427, (2019).



\bibitem{Capolupo2020Curv}
 A. Capolupo, G. Lambiase, A. Quaranta,
Phys. Rev. D, {\bf 101}, 095022 (2020).









\end{thebibliography}
\end{document}